\documentclass[aps,floatfix,prb,twoside,twocolumn,10pt,showpacs,citeautoscript,superscriptaddress]{revtex4-2}

\usepackage{amsmath}
\usepackage{amssymb}
\usepackage{booktabs}
\usepackage{float}
\usepackage{glossaries}
\usepackage{graphicx}
\usepackage{tabularx}
\usepackage{units}
\usepackage{xr-hyper}
\usepackage[dvipsnames]{xcolor}

\graphicspath{{./figures/}}

\usepackage[colorlinks=true]{hyperref}
\hypersetup{
    pdffitwindow=false,     % window fit to page when opened
    pdfstartview={FitH},    % fits the width of the page to the window
    pdfnewwindow=true,      % links in new PDF window
    colorlinks=true,
    linkcolor=NavyBlue,
    citecolor=NavyBlue,
    filecolor=NavyBlue,
    urlcolor=NavyBlue,
}

\newcommand{\phys}{
  Chalmers University of Technology,
  Department of Physics,
  Gothenburg, Sweden
}
\newcommand{\mc}{
  Chalmers University of Technology,
  Department of Microtechnology and Nanoscience,
  Gothenburg, Sweden
}

\AtBeginDocument{
\heavyrulewidth=.08em
\lightrulewidth=.05em
\cmidrulewidth=.03em
\belowrulesep=.65ex
\belowbottomsep=0pt
\aboverulesep=.4ex
\abovetopsep=0pt
\cmidrulesep=\doublerulesep
\cmidrulekern=.5em
\defaultaddspace=.5em
}

\setacronymstyle{long-short}
\newacronym{cb}{CB}{conduction band}
\newacronym{cbm}{CBM}{conduction band minimum}
\newacronym{cc}{CC}{configuration coordinate}
\newacronym{ctl}{CTL}{charge transition level}
\newacronym{dft}{DFT}{density functional theory}
\newacronym{dos}{DOS}{density of states}
\newacronym{homo}{HOMO}{highest occupied molecular orbital}
\newacronym{hr}{HR}{Huang-Rhys}
\newacronym{ipr}{IPR}{inverse participation ratio}
\newacronym{ks}{KS}{Kohn-Sham}
\newacronym{ld}{LD}{localized-to-delocalized}
\newacronym{ll}{LL}{localized-to-localized}
\newacronym{lo}{LO}{longitudinal optical}
\newacronym{lo-to}{LO-TO}{longitudinal optical-transverse optical}
\newacronym{lumo}{LUMO}{lowest unoccupied molecular orbital}
\newacronym{psb}{PSB}{phonon sideband}
\newacronym{si}{SI}{Supplementary Information}
\newacronym{spe}{SPE}{single-photon emitter}
\newacronym{stem}{STEM}{scanning transmission electron microscopy}
\newacronym{vb}{VB}{valence band}
\newacronym{vbm}{VBM}{valence band maximum}
\newacronym{xc}{XC}{exchange correlation}
\newacronym{zpl}{ZPL}{zero-phonon line}

\newcommand{\nbvn}{N$_\text{B}$-V$_{\text{N}}$}
\newcommand{\nbvnneu}{N$_\text{B}$-V$_{\text{N}}^0$}
\newcommand{\cbvn}{C$_\text{B}$-V$_{\text{N}}$}
\newcommand{\cbvnneu}{C$_\text{B}$-V$_{\text{N}}^0$}
\newcommand{\cbvnneg}{C$_\text{B}$-V$^{-}_{\text{N}}$}
\newcommand{\cbvnpos}{C$_\text{B}$-V$^{+}_{\text{N}}$}
\newcommand{\cnvb}{C$_\text{N}$-V$_{\text{B}}$}
\newcommand{\cnvbneu}{C$_\text{N}$-V$_{\text{B}}^0$}
\newcommand{\cnvbneg}{C$_\text{N}$-V$^{-}_{\text{B}}$}
\newcommand{\vn}{V$_{\text{N}}$}
\newcommand{\vnneu}{V$^0_{\text{N}}$}
\newcommand{\vnneuexc}{{V$^0_{\text{N}}$}$^{\ast}$}
\newcommand{\vb}{V$_{\text{B}}$}
\newcommand{\nb}{N$_{\text{B}}$}
\newcommand{\nbneu}{N$^0_{\text{B}}$}
\newcommand{\nbpos}{N$^{+}_{\text{B}}$}
\newcommand{\bn}{B$_{\text{N}}$}
\newcommand{\bnneu}{B$^0_{\text{N}}$}
\newcommand{\bnneg}{B$^{-}_{\text{N}}$}
\newcommand{\bnpos}{B$^{+}_{\text{N}}$}
\newcommand{\bnneuexc}{{B$^0_{\text{N}}$}$^{\ast}$}
\newcommand{\vbneu}{V$^0_{\text{B}}$}
\newcommand{\vbneg}{V$^{-}_{\text{B}}$}

\newcommand{\vbnegex}{{V$^{-}_{\text{B}}$}$^{\ast}$}
\newcommand{\cn}{C$_{\text{N}}$}
\newcommand{\cnneu}{C$^0_{\text{N}}$}
\newcommand{\cnneg}{C$^{-}_{\text{N}}$}
\newcommand{\cb}{C$_{\text{B}}$}
\newcommand{\cbneu}{C$^0_{\text{B}}$}
\newcommand{\cbpos}{C$^{+}_{\text{B}}$}
\newcommand{\vnneg}{V$^{-}_\mathrm{N}$}
\newcommand{\vnpos}{V$^{+}_\mathrm{N}$}
\newcommand{\nbvnneg}{N$_\mathrm{B}$-V$^{-}_\mathrm{N}$}
\newcommand{\nbvnpos}{N$_\mathrm{B}$-V$^{+}_\mathrm{N}$}
\newcommand{\qunit}{$\sqrt{\text{amu}}$\,{\AA}}

\newcommand{\cbcnneuexc}{C$_{\text{B}}$-C$^{0, \ast}_{\text{N}}$}
\newcommand{\cbcnneu}{C$_{\text{B}}$-C$^0_{\text{N}}$}

\newcommand{\spancol}[2]{\multicolumn{#1}{c}{#2}}

\begin{document}

\title{
    Vibrational signatures for the identification of \texorpdfstring{\\}{}
    single-photon emitters in hexagonal boron nitride
}
\author{Christopher Linder\"alv}
\email{chrlinde@chalmers.se}
\affiliation{\phys}
\author{Witlef Wieczorek}
\affiliation{\mc}
\author{Paul Erhart}
\email{erhart@chalmers.se}
\affiliation{\phys}

\begin{abstract}
Color centers in h-BN are among the brightest emission centers known yet the origins of these emission centers are not well understood.
Here, using first-principles calculations in combination with the generating function method, we systematically elucidate the coupling of specific defects to the vibrational degrees of freedom.
We show that the lineshape of many defects exhibits strong coupling to high frequency phonon modes and that C$_{\text{N}}$, C$_{\text{B}}$, \cb-\cn{} dimer and V$_{\text{B}}$ can be associated with experimental lineshapes.
Our detailed theoretical study serves as a guide to identify optically active defects in h-BN that can suit specific applications in photonic-based quantum technologies, such as single photon emitters, hybrid spin-photon interfaces, or spin-mechanics interfaces.
\end{abstract}

\maketitle

\section{Introduction}

Hexagonal boron nitride (h-BN) is a wide band gap van-der-Waals solid.
In its exfoliated form, h-BN is a stable two-dimensional material that retains its wide band gap of about \unit[6]{eV}.
This large band gap supports a diversity of optically active defect centers, exhibiting a wide range of emission energies between 1.2 and \unit[5.3]{eV} \cite{DuLiLin16, TraBraFor15}.
Recently, it was shown that mono and few-layer h-BN can host room temperature \glspl{spe} \cite{TraBraFor15}, which sparked enormous interest in the field of photonic-based quantum technologies \cite{obrien_photonic_2009, aharonovich_solid-state_2016}.
Follow-up experiments studied \glspl{spe} in h-BN in more detail \cite{TranKiaNgu18, TraBraFor15, ExaHopGro17, ShoJayCon16, TraElbTot16, WigSchPoz19,BomBec19} and, remarkably, such emitters were shown to exhibit Fourier transform-limited emission up to room temperature \cite{dietrich_solid_2019}.
An understanding of the optical properties of \glspl{spe} in h-BN would enable selecting specific defect centers to serve as single photon sources \cite{aharonovich_solid-state_2016}, as hybrid spin-photon \cite{atature_material_2018} or spin-mechanics interfaces \cite{abdi_spin-mechanical_2017, abdi_quantum_2019}.

In order to take full advantage of these \glspl{spe}, it is paramount to identify the responsible defect structures and corresponding electronic transitions.
However, so far no consensus has been reached concerning the origin of single-photon emission.
While the spatial localization of the single-photon emission strongly suggests point defects to be the culprit, the assignment to a specific defect in h-BN is hampered by the range of \glspl{zpl} that are observed and the large number of potential defect candidates.

For \glspl{zpl} between 1.6 and \unit[2.3]{eV}, in most cases pronounced \glspl{psb} at about \unit[165]{meV} below the \gls{zpl} are observed \cite{TranKiaNgu18, TraBraFor15, ExaHopGro17, ShoJayCon16, TraElbTot16, WigSchPoz19}.
In some experiments, \glspl{psb} for some \glspl{spe} exhibit, however, a double-peak structure at around \unit[160]{meV} and \unit[190-200]{meV} \cite{BomBec19, WigSchPoz19}.
The similarity of the \glspl{psb} across measurements suggests the emission to be due to either multiple different defects with similar geometry or a single defect with variable excitation energy.
In a recent experiment, the variable excitation energy was attributed to strain effects \cite{BomBec19}, but Stark shifts have been suggested as well \cite{XiaLiKim19}.
It has also been proposed that there are two families of emitters around \unit[2]{eV} with different electronic structure and \glspl{zpl} of 1.88 and \unit[2.14]{eV}, respectively, that can be distinguished not only based on their \gls{zpl} but also their quantum efficiency \cite{NikMenOze19}.

h-BN also exhibits luminescence in the ultraviolet \cite{MusFelKan08}, with a recent experiment \cite{BouMeuTar16} demonstrating single photon emission at \unit[4.1]{eV}.
The structure of this defect remains unknown but has been proposed to originate from carbon defects, although the emission intensity does not exhibit correlation with C impurity concentration \cite{TsuTsuUch18}.
The highest intensity peak of the \gls{psb} of the \unit[4.1]{eV} emitter has been suggested to originate from coupling to a \unit[187]{meV} local phonon mode \cite{MusFelKan08} and coupling to the zone center \gls{lo} phonon mode at \unit[200]{meV} \cite{VuoCasVal16}.

Recent theoretical studies focused on calculating the electronic structure of point defects in monolayer and bulk h-BN based on first-principle methods \cite{TawAliFro17, WesWicMac18}.
The \glspl{zpl} for rather many defects have been calculated within the accuracy permitted by current \gls{dft} methods based on hybrid functionals.
Furthermore, the \glspl{psb} have been analyzed using phenomenological models \cite{WigSchPoz19, FelPurLin19} using a few selected phonon modes.
To the best of our knowledge, a combined defect and \glspl{psb} study has, however, only been performed on the defects \nbvn{} and \cbvn{} \cite{TawAliFro17, GroMooCic20}.
In the latter studies it was concluded that the calculated emission spectra of \nbvn{} do not agree with the measured lineshapes but that \cbvn{} might be a \gls{spe}.
Besides these two defects, information about how specific defects couple to the vibrational degrees of freedom is scarce.

In the present work, we contribute to closing this knowledge gap concerning point defect-related emissions in h-BN by considering both charged and charge neutral transitions and the resulting emission lineshape for a set of the most common intrinsic and extrinsic point defects.
Importantly, we calculate the combined defect and \gls{psb} emission spectrum and, thus, can assess the vibrational fine structure.
The resulting emission spectrum can be used as an experimentally accessible fingerprint to identify defect-related \glspl{spe}.
This is possible since the vibrational fine structure of the emission spectrum due to an electronic transition on a point defect is highly sensitive to changes in the local distortion between initial and final state \cite{Mar59}.
In the following, we consider vacancies (\vn{}, \vb{}) and antisites (\nb{}, \bn{}) as the most important intrinsic defects as well as carbon impurities (\cb{}, \cn{}), vacancy-impurity complexes (\cbvn{}, \cnvb{}) and one antisite-vacancy complex (\nbvn{}).
This selection covers most of the defects that have been proposed as \glspl{spe} in h-BN.

\section{Methodology}

\subsection{Defect formation energy}
\label{sect:method-formation-energies}

The formation energy of a defect in charge state $q$ is given by
\begin{equation}
    \Delta E_f = E_\text{defect} - E_\text{ideal} -\sum \Delta N_i\mu_i + q(\varepsilon_\text{VBM} + \Delta \mu_e),
    \label{eq:eform}
\end{equation}
where $E_\text{defect}$ and $E_\text{ideal}$ are the total energies of the defective and ideal systems, respectively.
$\Delta N_i$ denotes the change in the number of atoms of type $i$ between defective and ideal system, while $\varepsilon_\text{VBM}$ and $\Delta \mu_e$ are the \gls{vbm} position and the (relative) electron chemical potential, respectively.
The chemical potentials $\mu_i$ of B and N are coupled to each other via
\begin{equation}
    \mu_\text{B} + \mu_\text{N} = -E_\text{BN},
\end{equation}
where $E_\text{BN}$ is the cohesive energy of h-BN.\footnote{
    We adhere to the convention that the cohesive energy is defined as the energy \emph{gained} upon formation from the atomic states and hence commonly a \emph{positive} quantity.
}
Below, we consider the nitrogen-rich limit, where $\mu_{\text{N}} = -\frac{1}{2}E_{\text{N$_2$}}$ and the boron-rich limit, where $\mu_\text{B}$ is taken as the negative cohesive energy of elemental $\alpha$-B (spacegroup R$\bar{3}$m).
The chemical potential of carbon is set to the one of graphene throughout.
The \gls{ctl} between charge states $q$ and $q'$ is the value of the electron chemical potential for which the formation energies of the defect in charge state $q$ and $q'$ are equal. Throughout this study, \glspl{ctl} are reported with respect to the \gls{vbm}.

\subsection{Lineshape of emission spectrum}
\label{sect:method-lineshape}

\Glspl{psb} arise due to emission from the electronic excited state to the vibrationally excited electronic ground state.
The structure of the emission spectrum can be computed from a knowledge of the phonon spectrum and the difference in the ionic configurations associated with excited and ground states $\Delta \boldsymbol{R}$.
The extent of the lattice distortion can be measured by the magnitude of the mass weighted difference in ionic displacements
\begin{equation}
    \Delta Q = \sqrt{\sum\limits_a m_a\Delta \boldsymbol{R}_a\cdot \Delta \boldsymbol{R}_a},
\end{equation}
where the sum runs over all atoms in the defect cell.

In this work, the lineshape is computed using the generating function approach \cite{Lax52, KubToy55, Mar59}.
The central quantity in the generating function approach is the electron-phonon spectral function, which depends on the coupling between lattice displacement and vibrational degrees of freedom.
The latter information is encoded in the so called \emph{partial} \gls{hr} factor
\begin{equation}
    S_k = \frac{1}{2\hbar} \omega_k Q_k^2.
\end{equation}
This expression is obtained in the low temperature and parallel mode approximation, i.e., the frequencies and eigenvectors of both ground and excited electronic states are related by a simple translation.
$Q_k$ is the projection of the lattice displacement on the normalized collective displacement described by phonon mode $k$ and given by \cite{AlkBucAws14}
\begin{equation}
    Q_k = \sum_a \sqrt{m_a} \left(\boldsymbol{n}^a_k\big|\Delta \boldsymbol{R}_a\right),
\end{equation}
where $a$ runs over all the atoms in the computational cell and $n_k$ is the normalized ionic displacement vector corresponding to phonon mode $k$.

The electron-phonon spectral function is then obtained by summation over all modes
\begin{equation}
    S(\omega) = \sum_k S_k \delta_{\omega,\omega_k}.
    \label{eq:S_omega}
\end{equation}
It has dimensions of inverse energy in the same units as $\omega$.
The integral over the electron-phonon spectral function gives the  (total) \gls{hr} factor of the transition.
The electron-phonon spectral function is then transformed to the time domain to obtain $S(t)=\int d\omega S(\omega)\exp(-i\omega t)$ and the generating function $G(t)$ is obtained by exponentiation of $S(t)$
\begin{equation*}
    G(t) = e^{S(t)}.
\end{equation*}
Fourier transformation of the generating function yields the lineshape function
\begin{equation} \label{eq:A}
    A(\omega_{eg} - \omega) = \frac{1}{2\pi}\int_{-\infty}^{\infty}dt\, G(t)e^{i(\omega_{eg} - \omega)t - \kappa|t|}.
\end{equation}
Here, $\kappa$ is a broadening parameter that governs the width of the \gls{zpl}.
It is necessary for numerical reasons and does not represent thermal broadening.
It should be chosen as small as possible to minimize its effect on the spectrum and as large as necessary top achieve a smooth representation of the spectral functions.
Its role is thus akin to the smearing width that is adopted when computing electronic and phonon densities of states.
Similarly, there is a correlation between the convergence with respect to smearing and the number of modes (i.e. the size of the supercell in the present case or the density of the Brillouin zone mesh when computing a density of states).
Here, we use a value of \unit[6]{meV}, which is chosen to achieve the balance described above (see \autoref{sfig:A-vs-kappa} in the \gls{si}).

Finally, the lineshape function is related to the luminescence intensity via
\begin{equation}
    L(\omega) = C \omega^3 A(\omega),
\end{equation}
where $C$ is a normalization constant chosen such that $\int d\omega\, L(\omega) = 1$.

The localization of a phonon mode can be measured using the \gls{ipr} \cite{AlkBucAws14}
\begin{equation}
    \mathrm{IPR}_k = 1 \left/ \sum_a \left(\boldsymbol{n}_{k}^a \cdot \boldsymbol{n}_{k}^a\right)^2 \right. ,
\end{equation}
which can assume values between 1 and $N$, where $N$ is the number of atoms in the computational cell for which the phonon spectrum has been calculated.
Smaller and larger values indicate more and less localized character.

\subsection{Computational details}

All \gls{dft} calculations were carried out using plane-wave basis sets \cite{KreFur96} and the projector augmented wave method \cite{Blo94, KreJou99} as implemented in the Vienna ab initio simulation package \cite{KreHaf93} (VASP, version 5.4.4).
Exchange-correlation contributions were obtained using the semi-local PBE functional \cite{PerBurErn96} and the hybrid HSE06 functional \cite{HeyScuErn03} using both the standard value for the mixing parameter $\alpha=0.25$ as well as a value of $\alpha=0.60$ tuned to reproduce the band gap as detailed below (\autoref{sect:results-ideal-material}).
A plane wave basis set with a cutoff energy of \unit[550]{eV} was employed to represent the electronic wave functions.
Geometry optimization was performed for all systems, during which the atomic positions were allowed to relax until all forces were less than \unit[20]{meV/\AA}.

Brillouin zone sampling was performed using a grid of $21 \times 21 \times 1$ for the primitive hexagonal (2-atom) unit cell.
Defect calculations were carried out using a $8\times 8\times 1$ supercell with \unit[19.88]{\AA} vacuum while the Brillouin zone was sampled using the $\Gamma$-point only.

The \glspl{zpl} arising from transitions between defect-induced levels and band states were computed using the HSE06 hybrid functional using a mixing parameter of $\alpha=0.6$ to correct for the band gap error of semi-local functionals (see \autoref{sect:results-ideal-material} for the motivation of this choice of mixing parameter).
For charged defect cells, a correction of $q^2\, \unit[0.5]{eV}$ was added to the total energy  to account for periodic image charge effects and potential offsets \cite{KomBerArk14, Komsa2020}.
This approach yields \glspl{ctl} that agree well (within \unit[30]{meV}) with the values presented in the erratum of Ref.~\citenum{KomBerArk14}.

Charge neutral excitations were modelled using the $\Delta$SCF method, in which the electronic occupations are constrained.
The description of electronic states that are localized at the level of semi-local exchange-correlation functionals such as PBE is commonly only weakly affected by the addition of exact exchange.
(For illustration, PBE and standard HSE06 ($\alpha=0.25$) yield \glspl{zpl} for $V_\text{B}^{-1}$ of \unit[1.71]{eV} and \unit[1.84]{eV}, respectively.)
Since in the present case the convergence of excited state calculations with hybrid functionals is both very tedious and computationally demanding, all charge neutral defect-defect transitions were treated at the semi-local \gls{dft} level (PBE).

Vibrational spectra were obtained at the semi-local \gls{dft} level (PBE) using the \textsc{phonopy} package \cite{TogTan15}.
In the computation of the electron phonon spectral function $S(\omega)$, the Kronecker $\delta$ in Eq.~\eqref{eq:S_omega} was approximated using normalized Gaussians with a smearing of \unit[6]{meV}.
The integration over the time domain in the Fourier transform to obtain $A(\omega)$ was performed over \unit[2]{ps} with a time spacing of \unit[1]{fs}.
Convergence of the spectral distribution function is demonstrated in the \gls{si}.

\section{Results}

\subsection{Pristine h-BN}
\label{sect:results-ideal-material}

Without taking into account zero-point effects, the PBE functional yields a lattice parameter of \unit[2.51]{\AA} for h-BN while one obtains \unit[2.49]{\AA} with HSE06 using $\alpha=0.25$.
These values compare well with the experimental lattice parameter for bulk BN of \unit[2.51]{\AA} at \unit[10]{K} \cite{PasPelKna02}.
For consistency, all calculations below, including those based on the standard HSE06 functional (mixing parameter $\alpha=0.25$), were carried out using a lattice constant of \unit[2.51]{\AA}.

The calculated \emph{electronic} (or single-particle) band gap, on the other hand, measures 4.67 and \unit[5.71]{eV} with PBE and standard HSE06 ($\alpha=0.25$), respectively, in line with other theoretical studies \cite{HaaStrPan18}.
The experimentally measured \emph{optical} band gap is \unit[6.1]{eV} and \unit[6.0]{eV} for monolayer h-BN \cite{ElaValPel19} and bulk BN \cite{CasValGil16}, respectively.
While these numbers appear rather close to the value obtained using standard HSE ($\alpha=0.25$), a sound comparison requires accounting for the exciton binding energy, which is very large both in bulk \cite{ArnLebRab06} and monolayer h-BN \cite{HaaStrPan18}, as well as the effect of zero-point motion via electron-phonon coupling \cite{TutMarWu18}.
It is therefore more instructive to compare the \gls{dft} values with the results from $G_0W_0$ calculations \cite{HaaStrPan18}, which provide a value of \unit[7.1]{eV}, corresponding to the single-particle band gap in the absence of both exciton formation and zero-point motion.
To reach this value using the HSE06 functional one requires a mixing parameter of $\alpha=0.60$, yielding a band gap of \unit[7.2]{eV}.
Below, we therefore use HSE06 with a mixing parameter of $\alpha = 0.60$ to compute defect formation energies and \glspl{zpl} derived thereof.

\begin{figure}
    \centering
    \includegraphics{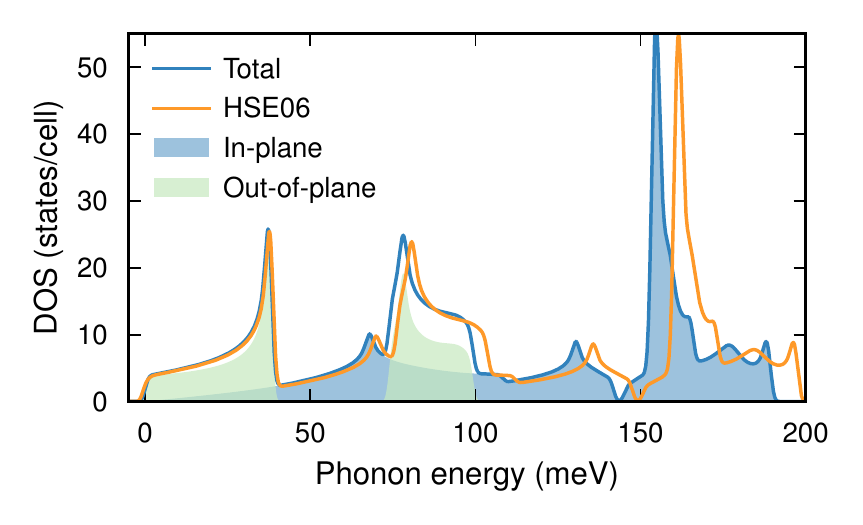}
    \caption{
        Vibrational spectrum of pristine h-BN with Cartesian projection of the normal modes as computed with PBE and HSE06 ($\alpha=0.25$ and total density of states only; computed using a $8\times 8\times 1$ supercell).
    }
    \label{fig:dos_ideal}
\end{figure}

Before considering the vibrational spectra of defects, a closer inspection of the vibrational spectrum of pristine h-BN is instructive.
The vibrational spectra from PBE and HSE06 are very similar, with the latter yielding a slightly stiffer response overall (\autoref{fig:dos_ideal}).\footnote{
  Since \gls{lo-to} splitting is absent in 2D materials \cite{SohGibCal17}, the non-analytical contribution to the force constant matrix has been omitted.
}
Given the 2D character of h-BN, the phonon \gls{dos} can be decomposed into an in-plane and out-of-plane part.
In part due to the quadratic dispersion inherent to 2D materials \cite{CarLiLin16}, the lower frequency part of the spectrum is dominated by out-of-plane modes.
The out-of-plane partial \gls{dos} features two pronounced bands ranging from 0 to about \unit[40]{meV} and from approximately 65 to \unit[100]{meV} (values from PBE), respectively, with two pronounced peaks at 40 and \unit[80]{meV}.
The in-plane partial \gls{dos} covers the entire frequency range spanning up to \unit[187]{meV} in good agreement with other first-principle studies \cite{SerBosAre07, TohKuwOba06}.
The most notable feature is an asymmetric peak at \unit[154]{meV}.

\subsection{Defect energetics}

\begin{figure*}
    \centering
    \includegraphics{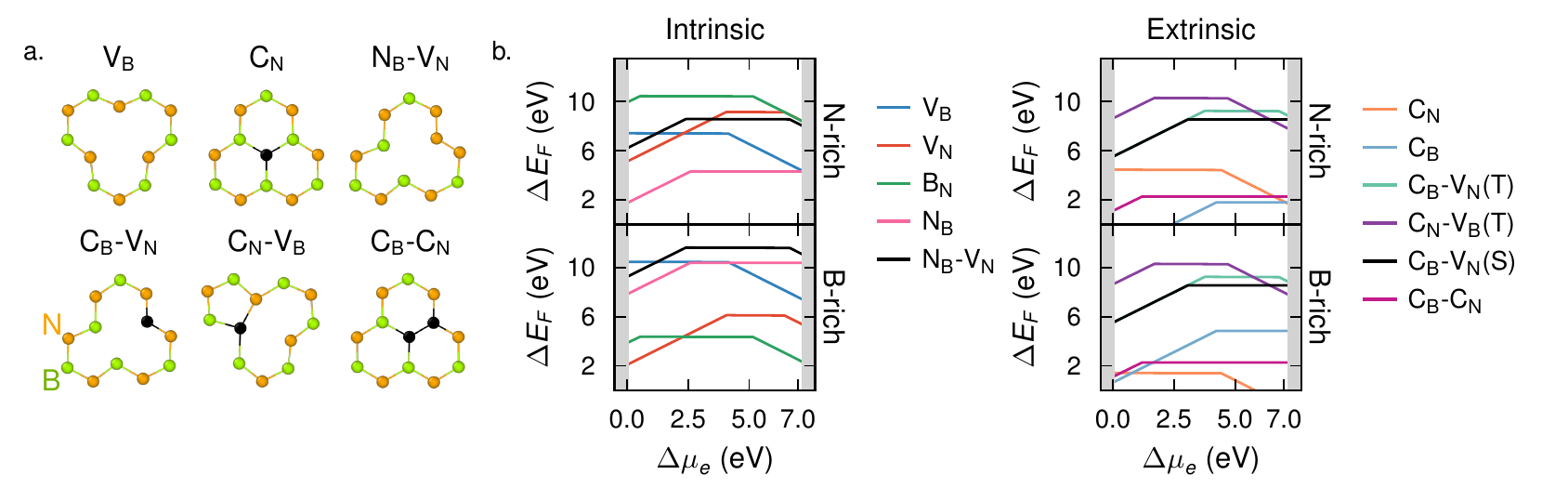}
    \caption{
        a) Structures of a selection of defects in h-BN.
        Boron, nitrogen, and carbon atoms are shown in green, orange, and black, respectively.
        b) Formation energies of intrinsic (left) and extrinsic (right) defects in the N-rich (top) and B-rich (bottom) limits.
        S and T indicate singlet and triplet configurations, respectively.
    }
    \label{fig:configurations_formation_energies}
\end{figure*}

From the outset we considered vacancies (\vn{}, \vb{}) and antisites (\nb{}, \bn{}) as possibly relevant intrinsic defects as well as carbon impurities (\cb{}, \cn{}, \cb{}-\cn{}), vacancy-impurity complexes (\cbvn{}, \cnvb{}) and one antisite-vacancy complex (\nbvn{}), see \autoref{fig:configurations_formation_energies}a for atomic structures.
To determine the energetically most favorable defects and defect configurations along with pertinent charge states we computed the defect formation energies under both B and N-rich conditions (\autoref{fig:configurations_formation_energies}b).
Where comparable the results are consistent with previous work on monolayer h-BN \cite{KomBerArk14}. %and bulk BN \cite{WesWicMac18}.
Under N-rich conditions \nb{} (donor with $q=+1$, 0) and \vb{} (acceptor with $q=0$, $-1$) are the most favorable intrinsic defects, whereas under B-rich conditions \bn{} (acceptor, $q=0$, $-1$) and \vn{} (ambipolar with $q=+1$, 0, $-1$) have the lowest formation energies.
With regard to extrinsic defects, we find \cb{} ($q=+1$, 0), \cn{} ($q=0$, $-1$), and \cb{}-\cn{} ($q=0$, $-1$) to be lowest energy defects under both B and N-rich conditions.
The defect complexes involving substitutional impurities and vacancies, specifically \cbvn{} and \nbvn{}, have high formation energies but are nonetheless included in the analysis below since they have been discussed as potential \glspl{spe} before \cite{TraBraFor15, TawAliFro17, ReiSajKob18}.

\subsection{Transition types}

In the following, we consider two types of transitions:
(i) \gls{ld} transitions involve a localized defect level and a (delocalized) band edge state;
(ii) \gls{ll} transitions involve two defect levels both of which reside inside the band gap when they are occupied.

To illustrate the features and emphasize similarities and differences between these transition types, it is instructive to recall the relations between configuration coordinate diagram, defect formation energies, \glspl{ctl}, and defect levels.

\begin{figure*}
    \centering
    \includegraphics{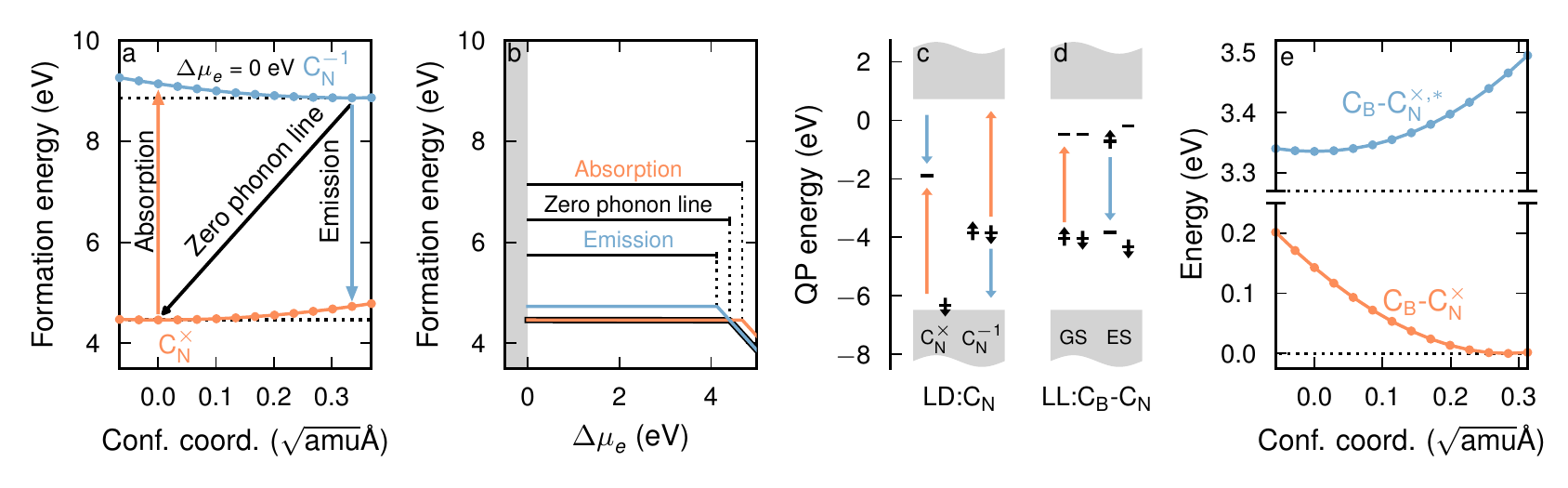}
    \caption{
      Relation between (a) transitions in the one-dimensional \gls{cc} diagram and (b) the corresponding formation energy diagram illustrated for the case of recombination with \gls{vb} charge carriers ($\Delta \mu_e = \unit[0]{eV}$) in {\cn}.
      The \gls{cc} diagram is constructed based on PBE values and shifted to the HSE06 ($\alpha=0.60$) \glspl{zpl}.
      The difference between localized-localized (LL) and localized-delocalized (LD) transitions is schematically shown in (c) and (d), where GS and ES stand for ground state and excited state, respectively, and the arrows indicate electrons in spin-up and spin-down states, respectively.
      In (e) the one-dimensional \gls{cc} diagram is shown for the \gls{ll} transition on {\cb}-{\cn}.
      The 1D \gls{cc} diagram has been computed with PBE.
    }
    \label{fig:configuration_coordinate_diagram}
\end{figure*}

In the case of \gls{ld} transitions, illustrated here by \cn{}, the absorption (emission) energy corresponds to the $0/-1$ \gls{ctl} obtained when the atomic configuration is constrained to the equilibrium \cnneu{} (\cnneg{}) geometry (\autoref{fig:configuration_coordinate_diagram}a).
The \gls{zpl} then corresponds to the \gls{ctl} obtained for the equilibrium geometries in either charge state (\autoref{fig:configuration_coordinate_diagram}b).
If the electron chemical potential $\Delta \mu_e$ coincides with the \gls{vbm}, emission occurs by capturing a hole from the \gls{vb} edge, depleting the defect level associated with \cn{} (\autoref{fig:configuration_coordinate_diagram}c).
If the electron chemical potential resides at the \gls{cbm} the initial and final states are inverted.
Since a delocalized band state is involved in this transition type, a proper description of the position of the band edges is crucial.

To illustrate \gls{ll} transitions, we consider here the \cb{}-\cn{} defect, which presents a particular simple two-level system.
Emission occurs from the excited state \cbcnneuexc{}, in which the highest occupied defect state resides above the lowest unoccupied level, to the ground state \cbcnneu{} (\autoref{fig:configuration_coordinate_diagram}d).
The difference in character of the highest occupied defect level in ground and excited states implies a difference in local potential that gives rise to a considerable lattice relaxation after emission (\autoref{fig:configuration_coordinate_diagram}e) that underlies the Stokes shift.

The possible \gls{ld} and \gls{ll} transitions involving the defects considered here are compiled in \autoref{tab:transitions}.

\subsection{Transitions on intrinsic defects}

Next we turn to a survey of the transition energies and subsequently an assessment of the lineshapes of intrinsic defects.
The prominent high-frequency \gls{psb} of emitters in the 1.6 to \unit[2.3]{eV} range and at \unit[4.1]{eV} indicates an effective phonon frequency of $\omega_{\text{eff}}\sim\unit[100]{meV}$.
The effective frequency is coupled to the \gls{hr} factor $S$ and the lattice distortion connecting the initial and final configurations $\Delta Q$ according to
\begin{align} \label{eq:hr}
    S = \omega_{\text{eff}} \Delta Q^2 \big/ 2\hbar.
\end{align}
As experimentally measured \gls{hr} factors fall in the range of $S=1$ to 2, the lattice distortion of potential emitters must therefore be relatively small.
For example, a transition coupled to an effective mode with a frequency of \unit[100]{meV} must have $\Delta Q<$\unit[0.41]{\qunit} in order to have an \gls{hr} factor below two.

Most transitions on intrinsic defects exhibit, however, rather large lattice distortions with $\Delta Q$ well above \unit[1]{\qunit} (see $\Delta Q$ in \autoref{tab:transitions}) that disqualify them as possible narrow band emitters.
These defects often exhibit geometries in which one or several atoms are located outside of the BN plane, which can be related to the large structural differences between different electronic states.
Transitions involving \vn{} and \nb{} are therefore not considered further.

\bn{} can host a \gls{ll} transition with a \gls{zpl} of \unit[1.14]{eV} according to PBE with a relatively small $\Delta Q$.
As the transition is rather far from the spectral range of interest here ($>\unit[1.6]{eV}$), transitions on \bn{} are not considered further either.

\nbvn{} can be ruled out as well based on the large value of $\Delta Q$.
The one remaining intrinsic defect \vb{} is considered in detail in the following section.

\subsubsection{Boron vacancy \texorpdfstring{(\vb)}{}}
\label{sect:results-boron-vacancy}

\begin{figure*}
    \centering
    \includegraphics{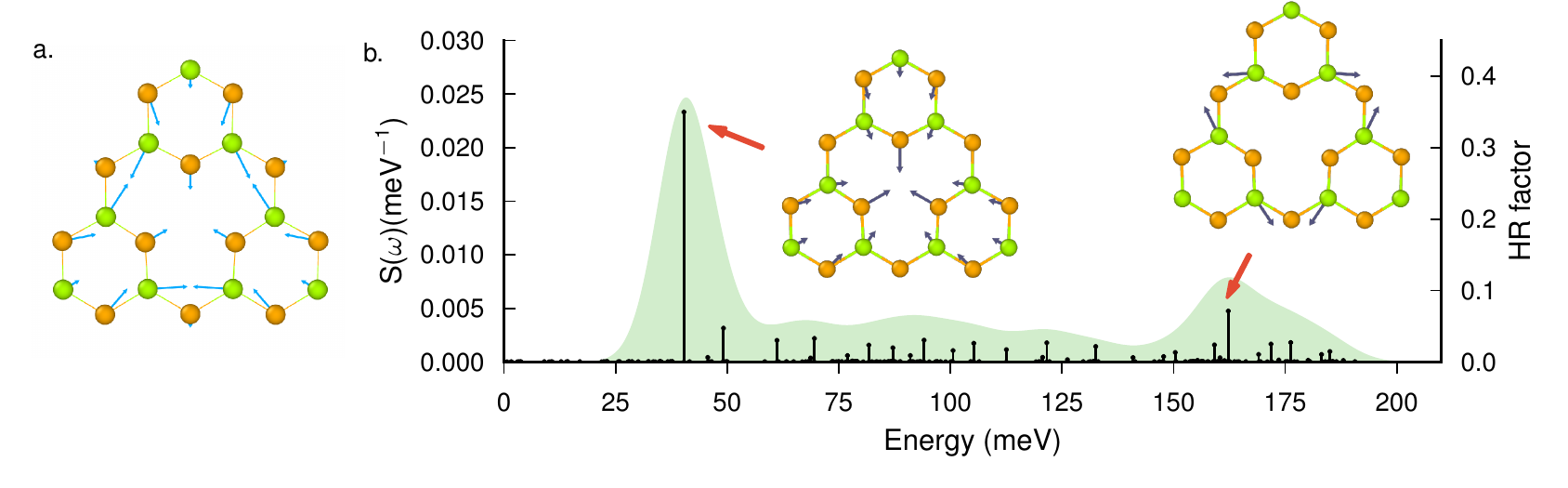}
    \caption{
        a) Ionic displacement in the $D_{3h}$ HOMO/LUMO transition on {\vbneg}.
        b) Spectral function and ionic displacements due to the dominant phonon modes.
    }
    \label{fig:displacement-boron-vacancy}
\end{figure*}

The symmetry of \vbneu{} is $C_{2v}$, while the symmetry of \vbneg{} is $D_{3h}$.
The $0/-1$ \gls{ctl} resides \unit[4.12]{eV} above the \gls{vbm}, which is much larger than an earlier theoretical value of \unit[2.4]{eV} \cite{AbdChoGal18}.
The latter though was obtained using a much smaller mixing parameter of $\alpha=0.32$, which as discussed in \autoref{sect:results-ideal-material} yields a misleading agreement with the experimentally measured \emph{optical} band gap.
The large difference is caused by a change in the spin state of {\vbneu} with increasing mixing parameter ($\alpha=0.25$: $S = 3/2$, $\alpha = 0.60$: $S=1/2$) that pushes the \gls{ctl} to higher energies with respect to the \gls{vbm} than the corresponding rigid band shift.

At the standard HSE06 level ($\alpha=0.25$), \vbneu{} induces 9 occupied in-gap states and 2 unoccupied in-gap states (see \autoref{sfig:kssta_f2} in the \gls{si} for the Kohn-Sham levels computed with standard HSE06).
In the spin-up channel, in which the optical transitions presumably occur, there are 5 occupied in-gap states and 2 degenerate unoccupied states.
There are two possible \gls{ld} transitions on \vb{}, namely a \gls{vb} hole capture transition involving \vbneg{} and a \gls{cb} electron capture by \vbneu{} with \glspl{zpl} of \unit[4.12]{eV} and \unit[3.09]{eV}, respectively.
The lattice distortion between \vbneu{} and \vbneg{} is quite large at \unit[1.0]{\qunit}.

Furthermore, there are 5 possible \gls{ll} transitions involving \vbneg{}.
Here, we consider only the \gls{homo}/\gls{lumo} transition in
both the $D_{3h}$ and ${C_{2v}}$ symmetries of the excited state with \glspl{zpl} of \unit[1.72]{eV} and \unit[1.63]{eV}, respectively.
The difference of $\sim$\unit[90]{meV} between the considered ionic configurations of the excited state suggests that ${C_{2v}}$ should be the equilibrium configuration.
Below we, however, consider the lineshapes of both the configurations with $D_{3h}$ and ${C_{2v}}$ symmetry, due to the small energy difference and the possibility that semi-local \gls{dft} provides an incorrect description of the equilibrium ionic configuration of the excited state.

First, the case where the initial state symmetry is $C_{2v}$ is considered with a total \gls{hr} factor of 2.45.
In this case there is a strong coupling of the electronic transition to a single mode at \unit[26]{meV} with a partial \gls{hr} factor of 0.66 (27\% of the total \gls{hr} factor) (the spectral function is shown in the \gls{si}, \autoref{sfig:spectral_function}).
The \gls{ipr} of this mode is 92, suggesting that it is likely a delocalized mode (the maximal value of the \gls{ipr} that can be reached in this supercell is 128).
At energies around \unit[40]{meV}, there are 3 modes with partial \gls{hr} factors of 0.11 to 0.29 with \glspl{ipr} of 40 to 50.
In the high-frequency end, there is a coupling to a \unit[162]{meV} mode with a partial \gls{hr} factor of 0.05 (2\% of the total \gls{hr} factor).
The resulting normalized emission lineshape is shown in \autoref{fig:vb-intensity}, where the spectrum has been broadened by tuning the damping parameter ($\kappa$) in \autoref{eq:A} (see Fig.~S11 for an illustration of how the spectral distribution function changes with increasing $\kappa$ in the case of {\cn}).

Next, the emission lineshape of the transition from the $D_{3h}$ symmetric initial state is considered (\autoref{fig:displacement-boron-vacancy}a), which has a much smaller total \gls{hr} of 0.91.
The spectral function (\autoref{fig:displacement-boron-vacancy}b) has two prominent peaks at \unit[40]{meV} and \unit[162]{meV}.
The \unit[40]{meV} peak results from the coupling to a single phonon mode and the coupling to other phonons with energy in the vicinity of \unit[40]{meV} is very weak.
The \gls{ipr} of the \unit[40]{meV} mode is 50 and the partial \gls{hr} of this mode is 0.34 (37\% of total \gls{hr}).
The phonon eigenvector for the \unit[40]{meV} and the \unit[162]{meV} mode is shown in \autoref{fig:displacement-boron-vacancy}b.
The \gls{ipr} of the \unit[162]{meV} mode is 26 and the partial \gls{hr} factor is 0.07.
While this value is slightly larger than the corresponding value in the emission from the $C_{2v}$ initial state, the relative contribution is much larger at 8\% of the total \gls{hr} factor.
Therefore, the peak at \unit[162]{meV} in the emission lineshape is much more pronounced in the case of emission from the $D_{3h}$ state.
The normalized emission lineshape of the emission from the $D_{3h}$ is shown in \autoref{fig:vb-intensity}.

\subsection{Transitions on extrinsic defects}

Compared with intrinsic defects among which only \vb{} is a viable candidate for single-photon emission, extrinsic defects involving carbon feature a multitude of suitable electronic transitions (\autoref{tab:transitions}).
Transitions involving {\cbvn}(S) and {\cnvb} exhibit large lattice distortions and are therefore excluded from further analysis as argued above.
In the following we therefore focus on {\cn}, {\cb}, {\cbvn}(T), as well as \cb-\cn{} dimers.

\begin{table*}
    \centering
    \caption{
        Transition energies and structural distortions associated with different transitions computed with PBE for localized-localized transitions and HSE06 for localized-delocalized transitions.
        HSE06 values were obtained by using HSE06 with $\alpha = 0.6$ on the standard HSE06($\alpha = 0.25$) ionic configuration.
        For localized-delocalized transitions both the \gls{zpl} corresponding to the transition involving the \gls{vb} edge (left column; identical to the charge transition level) and the transition involving the \gls{cb} edge (right column) are given (see \autoref{fig:configuration_coordinate_diagram}c).
        Arrows indicate the direction of a transition.
        Singlet and triplet states are marked by $(S)$ and $(T)$, respectively.
        Localized-delocalized transitions can proceed in either direction, whereas localized-localized transitions always commence from the excited state, the latter being marked by asterisks.
        The ``Modes'' column identifies which phonon spectrum was used for the computation of the lineshape function.
        ZPL: zero-phonon line;
        $\Delta Q$: magnitude of the lattice distortion between the two defect configurations involved in the transition;
        HR: Huang-Rhys factor.
    }
    \label{tab:transitions}
\begin{tabular}{llclrrcrrccc}

\toprule
Type
& \spancol{3}{Transition}
& \spancol{5}{ZPLs (eV)}
& $\Delta Q$ (\qunit) & Modes & HR factor \\
\cline{5-6}
\cline{8-9}
\\[-4pt]
& & & & \spancol{2}{PBE} && \spancol{2}{HSE06} & & & \\
\midrule
\\[-4pt]

Localized-delocalized
&  {\vbneu}             & $\longleftrightarrow{}$ & {\vbneg}           & 1.12   & 3.56      && 4.12 & 3.09  & 1.02  &                &      \\[3pt]
&  {\vnneu}             & $\longleftrightarrow{}$ & {\vnneg}           & 3.83   & 0.85      && 6.42 & 0.79  & 1.83  &                &      \\
&                       & $\longleftrightarrow{}$ & {\vnpos}           & 1.82   & 2.85      && 4.04 & 3.17  & 1.76  &                &      \\
&  {\nbneu}             & $\longleftrightarrow{}$ & {\nbpos}           & 0.58   & 4.08      && 2.56 & 4.65  & 1.71  &                &      \\[3pt]
&  {\bnneu}             & $\longleftrightarrow{}$ & {\bnneg}           & 2.89   &  1.79     && 5.14 & 2.07  & 1.21  &                &      \\
&  {\bnneu}             & $\longleftrightarrow{}$ & {\bnpos}           & --$^*$ & --$^*$    && 0.51 & 6.70  & 1.21  &                &      \\
&  {\nbvnneu}           & $\longleftrightarrow{}$ & {\nbvnneg}         & 3.83   & 0.85      && 6.63 & 0.58  & 2.10  &                &      \\
&                       & $\longleftrightarrow{}$ & {\nbvnpos}         & 0.73   & 3.95      && 2.38 & 4.83  & 1.62  &                &      \\[3pt]
\cline{2-12}
\\[-4pt]
&  {\cnneg}             & $\longleftrightarrow{}$ & {\cnneu}           & 2.23   & 2.45      && 4.40 & 2.81  & 0.37  & {\cnneu}       & 1.88 \\[3pt]
&  {\cbpos}             & $\longleftrightarrow{}$ & {\cbneu}           & 2.24   & 2.43      && 4.20 & 3.01  & 0.39  & {\cbpos}       & 1.80 \\[3pt]
&  {\cbvnneu}(S)        & $\longleftrightarrow{}$ & {\cbvnpos}(S)      & 0.95   & 3.72      && 3.02 & 4.20  & 4.19  &                &      \\
&                       & $\longleftrightarrow{}$ & {\cbvnneg}(S)      & 4.43   & 0.24      && 6.88 & 0.33  & 1.17  &                &      \\[3pt]
&  {\cbvnneu}(T)        & $\longleftrightarrow{}$ & {\cbvnpos}(T)      & 1.65   & 3.02      && 3.73 & 3.48  & 0.56  & {\cbvnneu}(T)  & 3.22 \\
&                       & $\longleftrightarrow{}$ & {\cbvnneg}(T)      & 4.22   & 0.45      && 6.78 & 0.43  & 0.60  & {\cbvnneu}(T)  & 3.31 \\[3pt]
&  {\cnvbneu}           & $\longleftrightarrow{}$ & {\cnvbneg}         & 2.44   & 2.23      && 4.67 & 2.54  & 3.14  &                &      \\[3pt]
&  {\cb}-{\cnneu}       & $\longleftrightarrow{}$ & {\cb}-{\cn}$^{+1}$ & --$^*$ & --$^*$    && 1.17 & 6.04  & 0.38  &                &      \\
\midrule

Localized-localized
& {\vbnegex} ($D_{3h}$) & $\longrightarrow{}$     & {\vbneg}           & \spancol{2}{1.72}  &&      &       & 0.34  & {\vbneg}       & 0.91 \\[3pt]
& {\vbnegex} ($C_{2v}$) & $\longrightarrow{}$     & {\vbneg}           & \spancol{2}{1.63}  &&      &       & 0.70  & {\vbneg}       & 2.45 \\[3pt]
& {\vnneuexc}           & $\longrightarrow{}$     & {\vnneu}           & \spancol{2}{0.74}  &&      &       & 2.10  &                &      \\[3pt]
& {\bnneuexc}           & $\longrightarrow{}$     & {\bnneu}           & \spancol{2}{1.14}  &&      &       & 0.75  &                &      \\[3pt]
\cline{2-12}
\\[-4pt]
& {\cb}-{\cn}$^{\ast}$  & $\longrightarrow{}$     & {\cb}-{\cnneu}     & \spancol{2}{4.07}  &&      &       & 0.28  & {\cb}-{\cnneu} & 1.10 \\[3pt]
& {\nbvn}$^{\ast}$      & $\longrightarrow{}$     & {\nbvnneu}         & \spancol{2}{1.98}  &&      &       & 1.06  &                &      \\[3pt]
& {\cbvn}$^{\ast}$(S)   & $\longrightarrow{}$     & {\cbvnneu}(S)      & \spancol{2}{1.04}  &&      &       & 2.91  &                &      \\
& {\cbvn}$^{\ast}$(T)   & $\longrightarrow{}$     & {\cbvnneu}(T)      & \spancol{2}{1.33}  &&      &       & 0.55  & {\cbvnneu}(T)  & 1.72 \\

\bottomrule
\multicolumn{9}{l}{$^*$ In the case of PBE, the \glspl{ctl} fall outside the band gap.}
\end{tabular}
\end{table*}

\subsubsection{Carbon-on-nitrogen \texorpdfstring{(\cn)}{}}

The ideal h-BN structure is only modified slightly with the inclusion of a {\cnneu}.
{\cnneu} in $D_{3h}$ symmetry exhibits a single unoccupied defect level within the band gap, indicating that only \gls{ld} transitions are possible (\autoref{fig:configuration_coordinate_diagram}c).
The $0/-1$ \gls{ctl} (\gls{zpl} for hole capture on {\cnneg}) resides at \unit[4.40]{eV}.
This is higher energy than earlier calculations for bulk h-BN that obtained a value of \unit[2.84]{eV} \cite{WesWicMac18}.
This can again be primarily attributed to the smaller mixing parameter of $\alpha=0.31$ (see \autoref{sect:results-ideal-material} and \autoref{sect:results-boron-vacancy}).
It is significantly larger than measurements that located the {\cn} acceptor at \unit[2.3]{eV} above the \gls{vbm} \cite{UddLiLin17}.
The other possible transition is via \gls{cb} electron capture on \cnneu{}, which has a \gls{zpl} of \unit[2.81]{eV}.
The lattice distortion associated with these transitions is only \unit[0.37]{\qunit}.

The transitions on {\cn} couple strongly to high frequency phonon modes, with the spectral function (\autoref{sfig:spectral_function}) exhibiting distinct peaks at \unit[158]{meV} and \unit[185]{meV}.
The \unit[158]{meV} peak consists of a single phonon mode that has a \gls{hr} factor of 0.23 (12.4\% of the total \gls{hr} factor of 1.88).
The \gls{ipr} of the \unit[158]{meV} mode is 27.
The \unit[185]{meV} peak on the other hand consists of several modes between \unit[182]{meV} and \unit[187]{meV}.
The largest contribution to the spectral function comes from one mode at \unit[182]{meV} with a \gls{hr} factor of 0.19 and an \gls{ipr} of 73, and two modes at \unit[187]{meV} with \gls{hr} factors of 0.19 and 0.22.
The \gls{ipr} of the \unit[187]{meV} modes are 75 and 73, respectively.
The normalized emission lineshape is shown in \autoref{fig:cn-cb-cn-intensity}a.

\begin{figure}
    \centering
    \includegraphics[scale = 0.2]{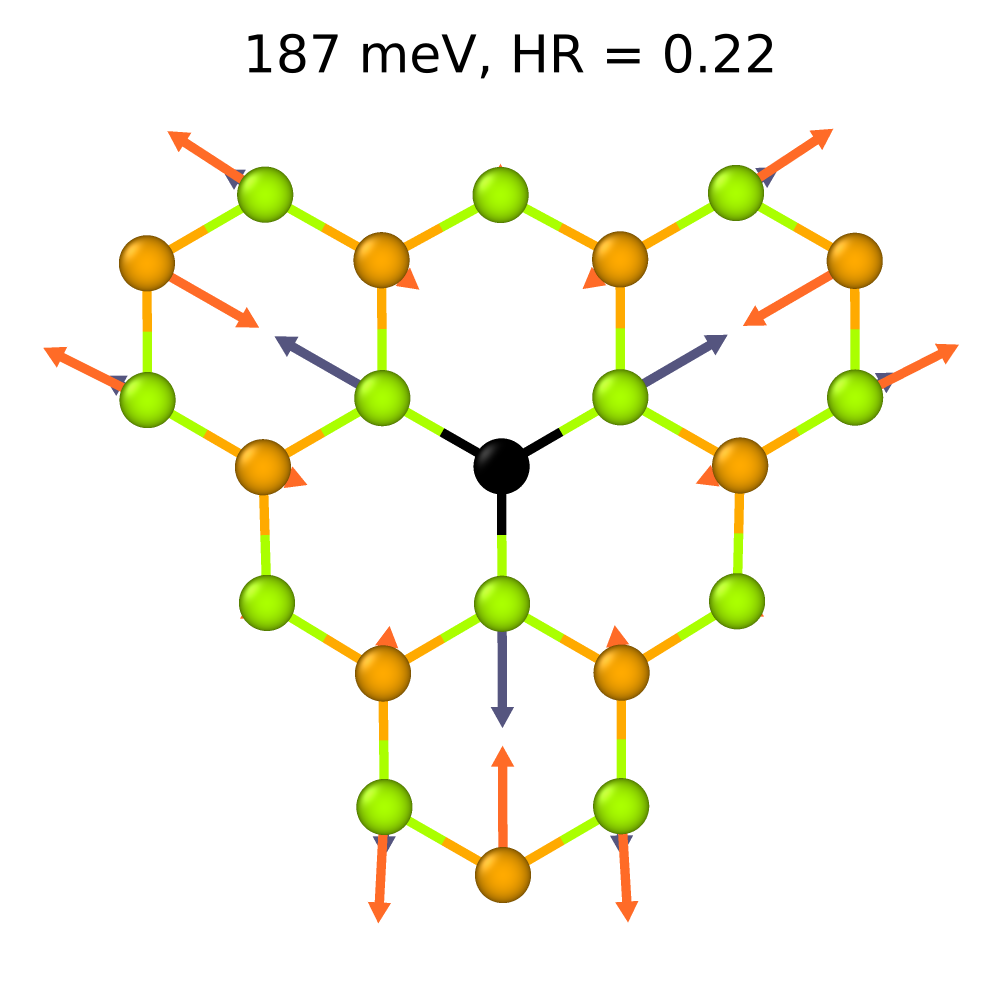}
    \caption{
        Phonon normal mode (red arrows) and transition displacements (dark arrows) for the \unit[187]{meV} mode coupling in the charged transition on {\cn}.
        The phonon normal mode is amplified by a factor of 15 and the transition displacements are amplified by a factor of 30.
    }
    \label{fig:cn-phonon-displacements}
\end{figure}

In order to elucidate the structural origin of the \gls{psb}, the ionic displacement due to the electronic transition can be overlaid with the phonon displacement vector for the highest frequency mode at \unit[187]{meV} (\autoref{fig:cn-phonon-displacements}).
The coupling between lattice and electronic transition in \cn{} is dominated by the displacement of B atoms.
The B ions closest to {\cnneg} experience the largest displacement upon transition to {\cnneu}.
However, these ions are not displaced in the phonon displacement vector so the contribution to the partial \gls{hr} is essentially zero.
The 6 B atoms in the next shell do not displace as much but contribute much more to the partial \gls{hr} factor due to a much larger overlap with the phonon eigenvector.
Finally, the B atoms in the third neighbour shell contribute the most to the partial \gls{hr} factor.

The vibrational coupling can also be approximated by the 1D \gls{cc} diagram in \autoref{fig:configuration_coordinate_diagram}a.
We find that the effective frequencies $\omega_{\text{eff}}$ determined from the potential energy surface are \unit[141]{meV} and \unit[146]{meV} for the ground state and excited state, respectively.
These frequencies translate into \gls{hr} factors (\autoref{eq:hr}) of 1.89 for the ground state and 1.95 for the excited state (1.88 with the generating function method).
The effective frequencies are determined by the coefficient $a$ in the fitted polynomial $aQ^2 + bQ^3$, from which we find $a/b > 1000$ suggesting that the harmonic approximation is sound.

\subsubsection{Carbon-on-boron \texorpdfstring{(\cb)}{}}

\begin{figure}
    \centering
    \includegraphics{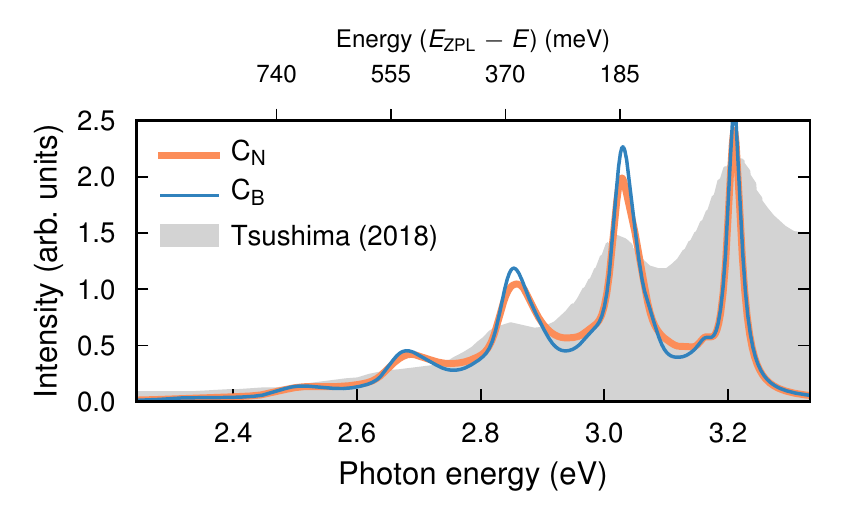}
    \caption{
        Emission lineshape of \cn{} and \cb{} defects compared with experimental results from Ref.~\citenum{TsuTsuUch18}.
        For the calculated spectra \glspl{zpl} has been shifted to the experimental value of \unit[3.22]{eV}.
        The experimental spectrum was measured at room temperature. The width of the \gls{zpl} has been broadened.
    }
    \label{fig:cn-cb-322}
\end{figure}

The possible charge transitions on {\cb} have \glspl{zpl} of \unit[4.20]{eV} and \unit[3.01]{eV}, and the structural distortion associated with these transitions is \unit[0.34]{\qunit}.
While the phonon spectrum for the {\cbneu} defect contains imaginary modes, we were unable to find lower energy structure by eigenmode following.
We attribute this finding to the very small energy difference associated with a displacement of the {\cb} atom perpendicular to the h-BN plane (see \autoref{sfig:energy_vs_c_displacement} of the \gls{si}).
The \gls{hr} factor is 1.80 and there is a well defined peak in the \glspl{psb} at \unit[185]{meV}, with additional phonon replicas at higher energies.
We note that the \glspl{zpl} and the spectral distribution function for \cb{} are very similar to the ones in \cn{} (\autoref{sfig:lineshapes}, \autoref{fig:cn-cb-322}), which would make it very difficult to distinguish between \cb{} and \cn{} in a photoluminescence experiment.

\subsubsection{Carbon-on-boron--nitrogen vacancy complex \texorpdfstring{(\cbvn)}{}}

The {\cbvn} defect is found to have a singlet ground state in agreement with previous studies \cite{ReiSajKob18, SajReiFor18} but as other studies have already pointed out a triplet electron configuration is also possible \cite{TawAliFro17, WuGalSun17}.
For the {\cbvn} we considered three different structures:
(\emph{i}) the triplet planar configuration (T),
(\emph{ii}) the singlet planar configuration (S-planar), and
(\emph{iii}) the singlet structure in which the C atom is displaced out-of-plane (S).

The singlet planar (S-planar) structure is dynamically unstable confirming previous reports \cite{NohChoKim18}.
The triplet structure (T), on the other hand, is dynamically stable but about \unit[0.2]{eV} higher in energy than the S-planar configuration with PBE (compare \autoref{sfig:cbvn_cc_diagram}).
Finally, the singlet out-of-plane structure, which is \unit[0.53]{eV} lower in energy than the S-planar configuration, is both thermodynamically and dynamically stable and thus should be the equilibrium configuration of the {\cbvn} defect.
(A careful comparison of singlet and triplet configurations, including an assessment of the role of the exchange-correlation functional can be found in the \gls{si}.)

While the singlet state is by far the most stable configuration, it can be ruled out as a \gls{spe} source due to the large value of $\Delta Q=$\unit[2.91]{\qunit}.
The \gls{ll} transition in the triplet state, on the other hand, exhibits a much smaller lattice distortion.
The \gls{zpl}, located at \unit[1.33]{eV}, has a large intensity and there are multiple peaks in the \gls{psb}, occurring at energies between \unit[28]{meV} and \unit[157]{meV} (\autoref{sfig:lineshapes}).
The high-energy peak is barely distinguishable from the spectral distribution function.
The \gls{hr} factor for this transition is 1.72, in good agreement with previous studies on the lineshape of {\cbvn} in
the triplet state \cite{TawAliFro17}.
The spectral distribution function for charged transitions on {\cbvn}(T) exhibit a pronounced intensity on the \gls{zpl} (see \autoref{sfig:spectral_function}). However, the \gls{psb} is wide and does not exhibit a shape that can be associated with experimental lineshapes.
The \gls{hr} factor for these transitions are 3.2 and 3.3.

\begin{figure*}
    \centering
    \includegraphics{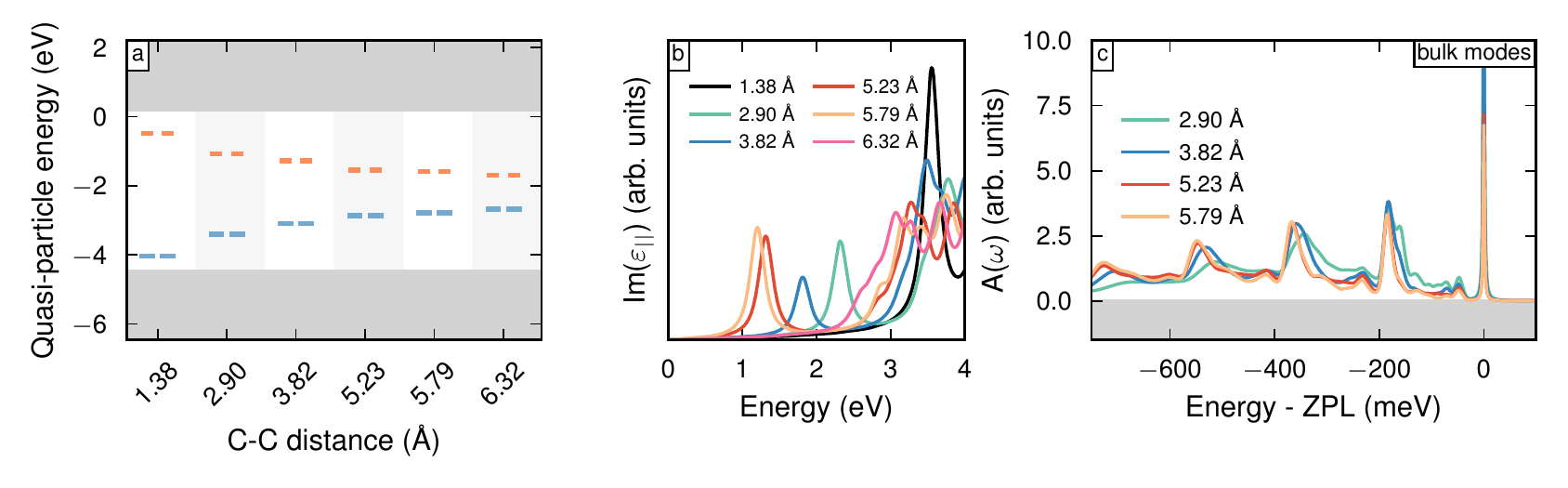}
    \caption{
        a) Kohn-Sham levels for {\cb}-{\cn} defect pairs as a function of separation from PBE calculations.
        b) Imaginary part of dielectric function (averaged in plane).
        c) Lineshape function for  {\cb}-{\cn} pairs.
    }
    \label{fig:cb-cn-pairs}
\end{figure*}

\subsubsection{Carbon-carbon dimer}

The \cb-\cn{} dimer consists of {\cn} and {\cb} defects located on neighboring lattice sites (\autoref{fig:configurations_formation_energies}a) and has been suggested to form at high C-doping levels \cite{UddLiLin17}.
The C-C bond is significantly shorter than the corresponding pristine B-N distance, and a mode with higher frequency than any pristine h-BN mode is present in the vibrational spectrum of the \cb-\cn{} dimer at \unit[195]{meV}.

The neutral charge state of the \cb-\cn{} dimer is thermodynamically stable for Fermi levels above \unit[1.17]{eV} making charged transitions in either the \unit[4.1]{eV} or \unit[1.6]{eV}--\unit[2.3]{eV} region possible (\autoref{fig:configurations_formation_energies}b).
There are four in-gap single particle levels, two of which are occupied and two unoccupied, making a \gls{ll} transition possible (\autoref{fig:configuration_coordinate_diagram}d).
The emission energy for the charged \gls{ll} transition on the \cb-\cn{} dimer is \unit[3.34]{eV} and dipole allowed (\autoref{fig:cb-cn-pairs}b) while the lattice distortion between the ground and excited state equilibrium configurations is \unit[0.28]{\qunit}.

The coupling of the charge neutral emission to the vibrational degrees of freedom has been analyzed with both the 1D \gls{cc} diagram (\autoref{fig:configuration_coordinate_diagram}e) and the generating function method.
From the 1D \gls{cc} we find that the effective frequencies are \unit[110]{meV} and \unit[127]{meV} for the ground and excited state with \gls{hr} factors based on the effective frequencies of 1.06 and 1.23, respectively.
In comparison to the 1D \gls{cc} for {\cn}, however, the third-order coefficient carries a much larger relative weight.
The ratio between the second and third-order coefficients is below 300 in both cases suggesting that anharmonic effects are not completely negligible.

The spectral function $S(\omega)$ features a significant peak corresponding to coupling to the \unit[195]{meV} mode (\autoref{sfig:spectral_function}), which correlates with a partial \gls{hr} factor of 0.44 to be compared with the total \gls{hr} factor of 1.10 as computed via the spectral function.
The \unit[195]{meV} mode is localized, as  indicated by the small value of the \gls{ipr} of 4.5, and corresponds to a stretching of the C-C bond.
The computed \gls{hr} factors of 1.10, 1.06 or 1.23, depending on method of calculation, agree well with the measured \gls{hr} factor for the \unit[4.1]{eV} luminescence of 1.3 \cite{MusFelKan08}.

\begin{figure}
    \centering
    \includegraphics{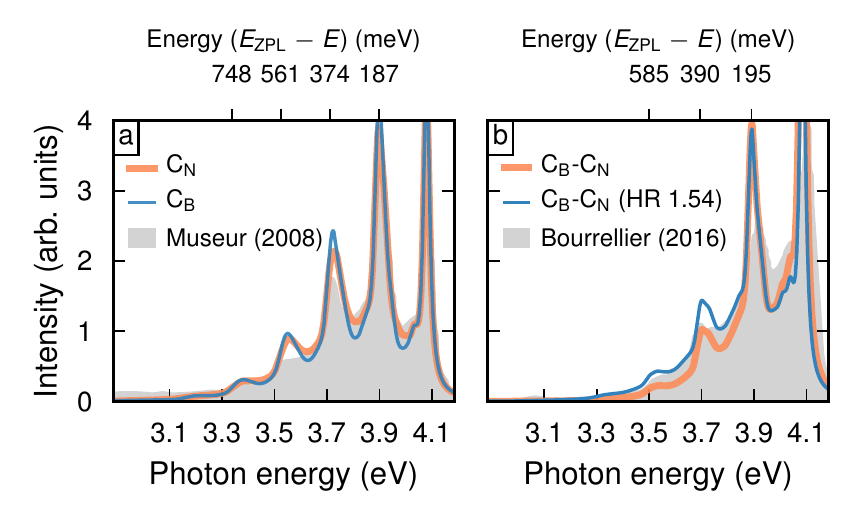}
    \caption{
        a) Calculated intensity for the $0$/$-1$ transition on {\cn} compared with experimental emission spectrum from Ref.~\citenum{MusFelKan08}.
        b) Calculated intensity for the charge neutral transition on {\cb}-{\cn} compared with measurements from Ref.\citenum{BouMeuTar16}.
        Measured and computed intensity are both normalized and the computed intensity has been shifted to the measured \gls{zpl}.
        The experimental spectra were measured at \unit[9]{K} (Ref.~\citenum{MusFelKan08}) and \unit[150]{K} (Ref.~\citenum{BouMeuTar16}), respectively. The width of the \gls{zpl} has been tuned to match experimental values.
    }
    \label{fig:cn-cb-cn-intensity}
\end{figure}

The computed normalized lineshape compares very well with experimentally  measured lineshapes \cite{BouMeuTar16} for the \unit[4.1]{eV} emission (\autoref{fig:cn-cb-cn-intensity}b), including both the positions of the features in the \glspl{psb} and the relative intensities between the first and second peaks.
We note that the results shown are for natural carbon i.e. $^{12}$C.
For $^{13}$C the frequency of the dominating mode at \unit[195]{meV} is reduced to \unit[191]{meV}.

\subsubsection{Dissociation of carbon-carbon dimer \texorpdfstring{(\cn-\cb)}{}}

Next, we consider the effect of spatial separation on the {\cn}-{\cb} defect, while the C-C distance is varied between \unit[2.90]{\AA} and \unit[6.32]{\AA}.
An inspection of the \gls{ks} levels shows that both \gls{ld} and \gls{ll} transitions are possible on the dissociated {\cn}-{\cb} pair (\autoref{fig:cb-cn-pairs}a).
The \gls{zpl} for the \gls{ld} $+1/0$ transition ranges from \unit[2.0]{eV} to \unit[2.7]{eV} (see \autoref{sfig:dissocated_cc_dimers}; all values given here include the band edge shift between PBE and HSE06 ($\alpha=0.6$)), while the \gls{ll} transitions exhibit \glspl{zpl} between \unit[2.39]{eV} at a separation of \unit[2.90]{\AA} and \unit[1.68]{eV} at a C-C separation of \unit[5.79]{\AA}, which is the longest C-C distance for which there is a well defined peak in the imaginary part of the dielectric function (\autoref{fig:cb-cn-pairs}b).

The vibrational properties of the dissociated {\cb}-{\cn} structures are computationally demanding to obtain due to the low symmetry of the systems.
Instead, coupling to bulk phonon modes is considered by utilizing the phonon eigenvectors for the ideal system.
Since the lattice distortion induced by \cb{} and \cn{} is small, the bulk phonon modes are expected to provide a good estimate of the emission lineshape.
The spectral distribution functions $S(\omega)$ for {\cn} computed with ideal modes and defective modes are very similar (see \autoref{sfig:cn_cbvn_lineshape}).
This is likely due to the limited structural relaxation relative to the ideal structure that occurs when incorporating C impurities in h-BN.
For defects such as {\cb}-{\vn}(T) (also shown in \autoref{sfig:cn_cbvn_lineshape}) that significantly distort the lattice, the spectral distribution computed with bulk modes is not a good approximation of the spectral distribution obtained using the modes of the defect structure.
This is especially true in cases where the contribution to the \gls{hr} factor mainly originates from coupling to local (small \gls{ipr}), defect induced modes.

The spectral distribution functions for the \gls{ll} transition on dissociated {\cb}-{\cn} pairs obtained in this fashion are shown in \autoref{fig:cb-cn-pairs}c for distances between \unit[2.91]{\AA} and \unit[5.79]{\AA}.
\cn{}-\cb{} defect pairs couple to high frequency bulk modes for all distances considered and have pronounced peaks separated by a frequency of around \unit[187]{meV} with some minor variations between the different structures.

\begin{table}
    \centering
    \caption{
        Transition energies and structural displacements from PBE calculations for charge neutral \gls{ll} transitions on \cn{}-\cb{} defect pairs.
    }
    \label{tab:cn-cb-pairs}
    \begin{tabular}{rrc}
       \toprule
       Distance ({\AA}) & ZPL (eV) & $\Delta Q$({\qunit}) \\
       \midrule
       1.38 & 3.34 & 0.28 \\
       2.90 & 2.39 & 0.42 \\
       3.82 & 2.03 & 0.44 \\
       5.23 & 1.75 & 0.47 \\
       5.79 & 1.68 & 0.46 \\
       6.32 & 1.57 & 0.48 \\
       \bottomrule
    \end{tabular}
\end{table}

\begin{figure}
    \centering
    \includegraphics{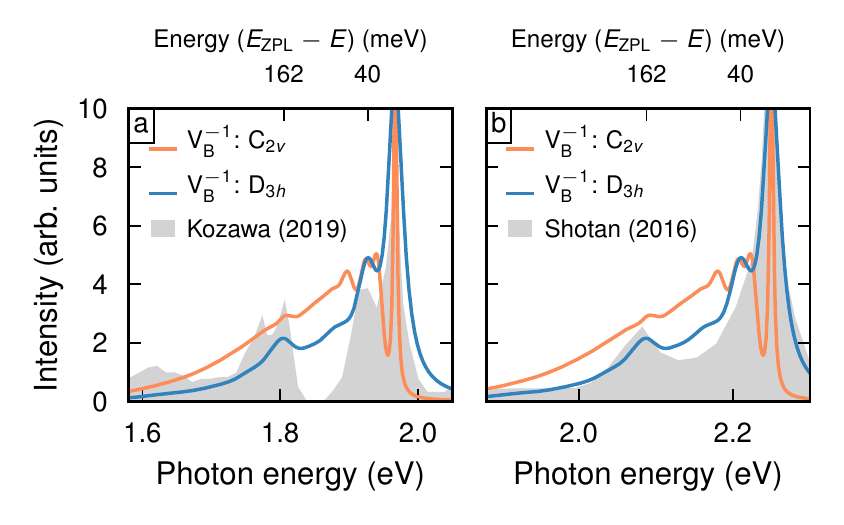}
    \caption{
        Calculated intensity for the charge neutral transition on {\vbnegex} compared with a) measurements from Ref.~\citenum{KozGovXin19} and b) measurements from Ref.~\citenum{ShoJayCon16}.
        The measured and computed intensity are both normalized and the computed intensity has been shifted to the measured \gls{zpl}.
        The experimental spectra in Ref.~\citenum{ShoJayCon16} were measured at room temperature. The width of the \gls{zpl} has been tuned to match experimental values.
    }
    \label{fig:vb-intensity}
\end{figure}

\section{Discussion}

The only intrinsic defect that emerges from our analysis of the vibrational spectra as a potential \gls{spe} (\vb{}) has a large formation energy.
Large formation energies are also obtained for many of the best candidates that are based on extrinsic defects.

The large formation energies are consistent with the observation that pristine h-BN usually exhibits a very small concentration of color centers.
To create emitters the h-BN flakes are usually subjected to some treatment, such as electron irradiation \cite{TraElbTot16}.
The experimental samples are therefore not equilibrated with respect to the environment, which means that defects with high formation energy may still be present and hence be sensible candidates for color centers.
This situation is facilitated by the strong bonding in h-BN, which gives rise to high activation energies for defect migration allowing the stabilization of high-energy defects over very long time scales.
As an extreme example, we note the recent preparation of large vacancy clusters \cite{KozGovXin19}, the vibrational signatures of which will be the subject of future work.

\subsection{Assignment of defects to the 1.6 to \texorpdfstring{\unit[2.3]{eV}}{2.3 eV} emitters}

Based on the computed \glspl{zpl} {\nbvn}, {\cbvn}, {\vbneg}, {\bn} and \cb{}-\cn{} pairs could be assigned to the \unit[2]{eV} emitters (\autoref{tab:transitions}).
\nbvn{} can be ruled out based on the magnitude of the relaxation between the states involved $\Delta Q$, which leads to an insensibly large \gls{hr} factor \cite{TraBraFor15, ShoJayCon16}.
\bn{} can be excluded on the same premises.
{\cbvn}(T), which has been proposed as a candidate \cite{TawAliFro17}, is an unlikely source due to the instability of the triplet state.
The more stable singlet state, on the other hand, is unlikely to exhibit a structured emission line based on the same argument as for the \nbvn{} defect.

\vbneg{} has been proposed to be an optically active defect and shown to be able to host transitions in the \unit[2]{eV} region before \cite{AbdChoGal18}.
The emission lineshape of the \gls{homo}/\gls{lumo} transition on {\vbneg} exhibits a rapidly decaying spectral weight away from the \gls{zpl} similar to many measured emission spectra in the \unit[2]{eV} band (\autoref{fig:vb-intensity}).
Very recently \gls{stem} images on vacancies and multi-vacancies in h-BN have become available \cite{KozGovXin19}.
{\vb} was identified and associated with an emission energy of \unit[1.98]{eV}, which is in very good agreement with \glspl{zpl} of \unit[1.6]{eV} to \unit[1.7]{eV} obtained from semi-local \gls{dft} calculations.
(We note that semi-local \gls{dft} calculations have been found to underestimate the \gls{ll} transition energy by around \unit[0.3]{eV} compared with HSE06 ($\alpha=0.25$) in the case of the NV$^{-1}$ center in diamond \cite{AlkBucAws14}
and the \cbvn(T) defect in monolayer h-BN \cite{TawAliFro17}.)

The computed lineshape for the \gls{ll} transition on {\vbneg} is compared in \autoref{fig:vb-intensity}a with the lineshape from Ref.~\citenum{KozGovXin19} that has been associated with \vb{}.
While some features agree between computed and experimental lineshapes such as the position of the first peak close to the \gls{zpl} and the peak at around \unit[162]{meV}, which is more pronounced in the case of emission from $D_{3h}$, the overall agreement is poor.
Specifically, the \gls{zpl} intensity is much larger in the computation and the region between the first peak and second peak carries a large spectral weight in the computation while there is a pronounced gap in most measurements.

While one could question whether semi-local \gls{dft} calculations are sufficient for modeling excited state geometries of {\vbneg}, the comparison with the \unit[2.25]{eV} emitter measured in Ref.~\citenum{ShoJayCon16} (\autoref{fig:vb-intensity}b) suggests that the lineshape observed in Ref.~\citenum{KozGovXin19} rather originates from some other defect.
Specifically, the computed lineshape for the {\vbneg} in $D_{3h}$ geometry is in good agreement with the measured spectrum, which supports the assignment of in-gap transitions on {\vbneg} to at least some emitters in the \unit[2]{eV} region.

The \glspl{zpl} for isolated carbon impurities (\cn{}, \cb{}) as well as the nearest-neighbor {\cb}-{\cn} dimer fall outside of the 1.6 to \unit[2.3]{eV} window considered here.
Next-nearest and farther neighbor {\cb}-{\cn} pairs (C-C distance $\geq\unit[2.90]{Å}$) exhibit, however, varying \glspl{zpl} around \unit[2]{eV} depending on separation distance (\autoref{tab:cn-cb-pairs}).
Unlike the (nearest-neighbor) \cb-\cn{} dimer, there are no direct C-C bonds present in these configurations and the \glspl{psb} mainly originate from coupling to bulk modes.
The lineshape exhibits only small changes with increasing C-C distance (\autoref{fig:cb-cn-pairs}) while the \gls{zpl} varies strongly (\autoref{tab:cn-cb-pairs}), which is a key feature of the \glspl{spe} found experimentally in the \unit[2]{eV} region.

The present analysis demonstrates that dissociated \cb-\cn{} defects correspond to a range of different \glspl{zpl} with very similar \glspl{psb}.
In practice, stabilization of these different \glspl{zpl} requires confinement of the atoms at specific lattice sites.
Since h-BN is a strongly covalent material bond breaking is energetically costly and migration barriers are high.
As a result, it is plausible that \cb-\cn{} defects can be stabilized at a range of distances.

\subsection{Assignment of defects to the \texorpdfstring{\unit[4.1]{eV}}{4.1 eV} emitter}

\subsubsection{Isolated carbon impurities \texorpdfstring{(\cb{} and \cn{})}{}}

The \unit[4.1]{eV} emission line has previously been suggested to originate from {\cn} \cite{BouMeuTar16} and carbon-carbon dimers (\cb{}-cn{}, \autoref{sect:discussion-c-c-dimer}) \cite{MacMacWal19}.
Focusing first on \cn{}, we find that it can host \gls{ld} transitions with \glspl{zpl} of \unit[4.4]{eV} (\gls{vbm} hole capture on {\cnneg}) and \unit[2.81]{eV} (\gls{cbm} electron capture on {\cnneu}), and moreover demonstrate that \cb{} exhibits similar \glspl{zpl}.
These \glspl{zpl} are in rather good agreement with the \unit[4.1]{eV} line.

The computed lineshapes for \cn{} and \cb{} show excellent agreement with the lineshape of a \unit[4.1]{eV} emitter reported in Ref.~\citenum{MusFelKan08} (\autoref{fig:cn-cb-cn-intensity}a).
The frequencies of the dominant modes in {\cn} and {\cb} are \unit[182]{meV} and \unit[187]{meV}, in excellent agreement with the measured frequencies.
It is, however, important to note that neither defect distorts the lattice significantly enough to induce local modes, which was argued in Ref.~\citenum{MusFelKan08} to be the origin of the \gls{psb}.
We note that the excellent agreement might be partly coincidental since in bulk samples, such as the one measured in Ref.~\citenum{MusFelKan08}, the \gls{lo-to} splitting is significant and might push the \gls{psb} away from the \gls{zpl} by $\sim\,\unit[10]{meV}$.
In fact, the highest intensity peak in the \gls{psb} for the \unit[4.1]{eV} emitter has been suggested to originate from coupling to the zone center \gls{lo} mode at \unit[200]{meV} \cite{VuoCasVal16}.
The present calculations for \cn{} and \cb{} yield lineshapes and \glspl{zpl} consistent with the \unit[4.1]{eV} emitter observed experimentally.

Our calculations also suggest \cb{} and \cn{} to be strong candidates for emitters found at 3.2 and \unit[3.4]{eV} \cite{BerKorTri16, TsuTsuUch18}.
The lineshapes of both \cn{} and \cb{} compare well with the \unit[3.22]{eV} emitter in Ref.~\citenum{TsuTsuUch18}, where the first \gls{psb} was found at \unit[200]{meV} from the \gls{zpl} with two additional distinct phonon replicas (\autoref{fig:cn-cb-322}).
The experimental data was recorded at room temperature, which explains the considerably broader spectrum compared to the calculation.
The position of the first \gls{psb} and the phonon replicas agree well between experiment and calculation safe for a $\sim\,\unit[10]{meV}$ offset, which, as noted above, arises from the presence of \gls{lo-to} splitting in a bulk sample.
Both \glspl{zpl} and lineshape thus suggest that \cn{} and \cb{} can be associated with an \unit[3.2]{eV} emitter.

\subsubsection{Carbon-carbon dimer \texorpdfstring{(\cb{}-\cn{})}{}}
\label{sect:discussion-c-c-dimer}

As noted above, the \unit[4.1]{eV} emission has also been associated with the (nearest-neighbor) carbon-carbon dimer (\cb{}-\cn{}) \cite{MacMacWal19}.
While \gls{dft} calculations based on semi-local \gls{xc} functionals yield a \gls{zpl} of \unit[3.34]{eV} for this defect, hybrid functionals with a larger fraction of exact exchange predict a \gls{zpl} in much better agreement with the measured \unit[4.1]{eV} emission line \cite{MacMacWal19}.
Here, it is important to keep in mind that the \gls{ll} transition involving \cb{}-\cn{} should be much less affected by the band gap error from the \gls{ld} transitions on isolated \cn{} or \cb{} since it only involves localized states, which are already well described by semi-local \gls{xc} functionals.

Our analysis shows that the lineshape of the \cb-\cn{} defect agrees very well with a measured emission spectrum from Ref.~\citenum{BouMeuTar16} (\autoref{fig:cn-cb-cn-intensity}b).
The \glspl{psb} are located at approximately the correct positions, although the spectral weight of the \gls{zpl} is slightly larger for the computed lineshape.
Interestingly, the \unit[195]{meV} mode that is the origin of the \gls{psb} in the \cb{}-\cn{} dimer is indeed a local mode in contrast to the dominant modes in {\cn} and {\cb}, which are bulk-like.
Furthermore, the computed \gls{hr} factor of 1.1 to 1.2 compares favorably with a measured value of 1.3 for the \unit[4.1]{eV} emitter \cite{MusFelKan08}.

The Stokes shift on the excited state landscape of \unit[0.2]{eV} obtained with HSE06 ($\alpha=0.25$) is notably larger than the value of \unit[0.13]{eV} from PBE.
A larger Stokes shift would indicate a smaller spectral weight of the \gls{zpl}.
Assuming that this difference is dominated by the displacement of the potential energy surfaces relative to each other, the spectral function can be renormalized and, in fact, renormalizing the electron-phonon spectral function to a \gls{hr} factor of 1.54 leads to an even better agreement.

We note that since the \gls{psb} originates to a large extent from the local mode at \unit[195]{meV}, which predominantly involves C motion, isotopic effects may appear in the \emph{position} of the \gls{psb}.
This is in contrast to, e.g., the transitions on {\cn} and {\cb}, where isotopic control over the C atoms in the host matrix should not affect the \gls{psb}.
Hence if one can isotopically control the formation of the {\cb}-{\cn} dimer to comprise a pair of $^{13}$C atoms instead of the naturally occurring $^{12}$C one should detect a small variation in the position of the \gls{psb} if the \unit[4.1]{eV} line originates from the {\cb}-{\cn} dimer.
Based on our calculations we estimate this frequency shift to be \unit[4]{meV}, placing the local C-C mode at \unit[191]{meV} for $^{13}$C.
While emitters based on $^{13}$C have been fabricated \cite{PelEliPag19}, changes in the \gls{psb} have not been investigated yet.

The transitions on intrinsic defects that have \glspl{zpl} in the vicinity of \unit[4.1]{eV} include hole capture on {\vnneu}, as well as hole capture on {\vbneg}, which give rise to \glspl{zpl} of \unit[4.04]{eV} and \unit[4.12]{eV}, respectively (see \autoref{tab:transitions}).
However, the structural distortions associated with these charged (\gls{ld}) transitions are relatively large and unlikely to be compatible with the measured lineshape and \gls{hr} factor (\autoref{tab:transitions}).

\section{Conclusions}

We have analyzed the optical transitions occurring from charge transitions (\gls{ld}) and charge neutral (\gls{ll}) transitions for a wide range of defects in h-BN.
For a selection of these defects, we have examined the possibility of assigning defect emission to well known \glspl{spe} by calculating line shapes and emission spectra with the generating function method.
We have found that {\vb}, {\cn}, {\cb}, and {\cb}-{\cn} can host transitions that couple strongly to high-frequency modes while still exhibiting a moderate \gls{hr} factor between 0.9 and 2.5.

The main conclusions are
\begin{itemize}
\item[(\emph{i})]
    {\vb}$^{-1}$ is likely to be a \gls{spe} with a narrow emission band with a \gls{zpl} in the \unit[1.6]{eV} to \unit[2.3]{eV} region. The lineshape shows decent agreement with the measured lineshape of an emitter with a \unit[2.25]{eV} \gls{zpl}.
\item[(\emph{ii})]
    The lineshapes of {\cn} and {\cb} are in excellent agreement with measured lineshapes for the \unit[4.1]{eV} emitter.
    Furthermore, the \glspl{zpl} are in good agreement with the \unit[4.1]{eV} \gls{zpl}.
    In addition, we note that these defects also exhibit \glspl{zpl} for the reverse transition that agrees well with the observed 3.2 to \unit[3.4]{eV} luminescence.
\item[(\emph{iii})]
    Next-nearest and farther neighbor (dissociated) {\cb}-{\cn} defect pairs allow for both charge neutral and charged transitions over a wide range of \gls{zpl} energies with narrow emission bands.
\item[(\emph{iv})]
    Our findings support the assignment of the \unit[4.1]{eV} emission to the \cb-\cn{} dimer (nearest-neighbor pair) based on the lineshape but we note that there are additional defects that exhibit similar emission lineshapes.
\end{itemize}

The current results corroborate the emerging consensus that the low frequency band (1.6 to \unit[2.3]{eV}) originates from {\vb} based defects and the high frequency ($\sim\!\unit[4.1]{eV}$) emitter originates from C based defects.
We note that recently additional defect configurations have been suggested that have not been considered here, including Stone-Wales defects \cite{HamThiBod20} as well as oxygen-carbon pairs \cite{VokWei20}.

\section*{Acknowledgements}
We are grateful to Arsalan Hashemi and Hannu-Pekka Komsa for fruitful discussions and for providing us with the finite-size corrections for the calculations of the defect formation energies in Ref.~\onlinecite{KomBerArk14}.
We acknowledge the Knut and Alice Wallenberg Foundation (2014.0226), the Swedish Research Council (2018-06482), and the Chalmers Excellence Initiative Nano for financial support.
The computations were enabled by resources provided by the Swedish National Infrastructure for Computing (SNIC) at NSC, C3SE and PDC partially funded by the Swedish Research Council through grant agreement no. 2018-05973.


\begin{thebibliography}{61}%
\makeatletter
\providecommand \@ifxundefined [1]{%
 \@ifx{#1\undefined}
}%
\providecommand \@ifnum [1]{%
 \ifnum #1\expandafter \@firstoftwo
 \else \expandafter \@secondoftwo
 \fi
}%
\providecommand \@ifx [1]{%
 \ifx #1\expandafter \@firstoftwo
 \else \expandafter \@secondoftwo
 \fi
}%
\providecommand \natexlab [1]{#1}%
\providecommand \enquote  [1]{``#1''}%
\providecommand \bibnamefont  [1]{#1}%
\providecommand \bibfnamefont [1]{#1}%
\providecommand \citenamefont [1]{#1}%
\providecommand \href@noop [0]{\@secondoftwo}%
\providecommand \href [0]{\begingroup \@sanitize@url \@href}%
\providecommand \@href[1]{\@@startlink{#1}\@@href}%
\providecommand \@@href[1]{\endgroup#1\@@endlink}%
\providecommand \@sanitize@url [0]{\catcode `\\12\catcode `\$12\catcode
  `\&12\catcode `\#12\catcode `\^12\catcode `\_12\catcode `\%12\relax}%
\providecommand \@@startlink[1]{}%
\providecommand \@@endlink[0]{}%
\providecommand \url  [0]{\begingroup\@sanitize@url \@url }%
\providecommand \@url [1]{\endgroup\@href {#1}{\urlprefix }}%
\providecommand \urlprefix  [0]{URL }%
\providecommand \Eprint [0]{\href }%
\providecommand \doibase [0]{http://dx.doi.org/}%
\providecommand \selectlanguage [0]{\@gobble}%
\providecommand \bibinfo  [0]{\@secondoftwo}%
\providecommand \bibfield  [0]{\@secondoftwo}%
\providecommand \translation [1]{[#1]}%
\providecommand \BibitemOpen [0]{}%
\providecommand \bibitemStop [0]{}%
\providecommand \bibitemNoStop [0]{.\EOS\space}%
\providecommand \EOS [0]{\spacefactor3000\relax}%
\providecommand \BibitemShut  [1]{\csname bibitem#1\endcsname}%
\let\auto@bib@innerbib\@empty
%</preamble>
\bibitem [{\citenamefont {Du}\ \emph {et~al.}(2016)\citenamefont {Du},
  \citenamefont {Li}, \citenamefont {Lin},\ and\ \citenamefont
  {Jiang}}]{DuLiLin16}%
  \BibitemOpen
  \bibfield  {author} {\bibinfo {author} {\bibfnamefont {X.~Z.}\ \bibnamefont
  {Du}}, \bibinfo {author} {\bibfnamefont {J.}~\bibnamefont {Li}}, \bibinfo
  {author} {\bibfnamefont {J.~Y.}\ \bibnamefont {Lin}}, \ and\ \bibinfo
  {author} {\bibfnamefont {H.~X.}\ \bibnamefont {Jiang}},\ }\href {\doibase
  10.1063/1.4941540} {\bibfield  {journal} {\bibinfo  {journal} {Applied
  Physics Letters}\ }\textbf {\bibinfo {volume} {108}},\ \bibinfo {pages}
  {052106} (\bibinfo {year} {2016})}\BibitemShut {NoStop}%
\bibitem [{\citenamefont {Tran}\ \emph {et~al.}(2015)\citenamefont {Tran},
  \citenamefont {Bray}, \citenamefont {Ford}, \citenamefont {Toth},\ and\
  \citenamefont {Aharonovich}}]{TraBraFor15}%
  \BibitemOpen
  \bibfield  {author} {\bibinfo {author} {\bibfnamefont {T.~T.}\ \bibnamefont
  {Tran}}, \bibinfo {author} {\bibfnamefont {K.}~\bibnamefont {Bray}}, \bibinfo
  {author} {\bibfnamefont {M.~J.}\ \bibnamefont {Ford}}, \bibinfo {author}
  {\bibfnamefont {M.}~\bibnamefont {Toth}}, \ and\ \bibinfo {author}
  {\bibfnamefont {I.}~\bibnamefont {Aharonovich}},\ }\href {\doibase
  10.1038/nnano.2015.242} {\bibfield  {journal} {\bibinfo  {journal} {Nature
  Nanotechnology}\ }\textbf {\bibinfo {volume} {11}},\ \bibinfo {pages} {37}
  (\bibinfo {year} {2015})}\BibitemShut {NoStop}%
\bibitem [{\citenamefont {O'Brien}\ \emph {et~al.}(2009)\citenamefont
  {O'Brien}, \citenamefont {Furusawa},\ and\ \citenamefont
  {Vučković}}]{obrien_photonic_2009}%
  \BibitemOpen
  \bibfield  {author} {\bibinfo {author} {\bibfnamefont {J.~L.}\ \bibnamefont
  {O'Brien}}, \bibinfo {author} {\bibfnamefont {A.}~\bibnamefont {Furusawa}}, \
  and\ \bibinfo {author} {\bibfnamefont {J.}~\bibnamefont {Vučković}},\
  }\href {\doibase 10.1038/nphoton.2009.229} {\bibfield  {journal} {\bibinfo
  {journal} {Nature Photonics}\ }\textbf {\bibinfo {volume} {3}},\ \bibinfo
  {pages} {687} (\bibinfo {year} {2009})}\BibitemShut {NoStop}%
\bibitem [{\citenamefont {Aharonovich}\ \emph {et~al.}(2016)\citenamefont
  {Aharonovich}, \citenamefont {Englund},\ and\ \citenamefont
  {Toth}}]{aharonovich_solid-state_2016}%
  \BibitemOpen
  \bibfield  {author} {\bibinfo {author} {\bibfnamefont {I.}~\bibnamefont
  {Aharonovich}}, \bibinfo {author} {\bibfnamefont {D.}~\bibnamefont
  {Englund}}, \ and\ \bibinfo {author} {\bibfnamefont {M.}~\bibnamefont
  {Toth}},\ }\href {\doibase 10.1038/nphoton.2016.186} {\bibfield  {journal}
  {\bibinfo  {journal} {Nature Photonics}\ }\textbf {\bibinfo {volume} {10}},\
  \bibinfo {pages} {631} (\bibinfo {year} {2016})}\BibitemShut {NoStop}%
\bibitem [{\citenamefont {Tran}\ \emph {et~al.}(2018)\citenamefont {Tran},
  \citenamefont {Kianinia}, \citenamefont {Nguyen}, \citenamefont {Kim},
  \citenamefont {Xu}, \citenamefont {Kubanek}, \citenamefont {Toth},\ and\
  \citenamefont {Aharonovich}}]{TranKiaNgu18}%
  \BibitemOpen
  \bibfield  {author} {\bibinfo {author} {\bibfnamefont {T.~T.}\ \bibnamefont
  {Tran}}, \bibinfo {author} {\bibfnamefont {M.}~\bibnamefont {Kianinia}},
  \bibinfo {author} {\bibfnamefont {M.}~\bibnamefont {Nguyen}}, \bibinfo
  {author} {\bibfnamefont {S.}~\bibnamefont {Kim}}, \bibinfo {author}
  {\bibfnamefont {Z.-Q.}\ \bibnamefont {Xu}}, \bibinfo {author} {\bibfnamefont
  {A.}~\bibnamefont {Kubanek}}, \bibinfo {author} {\bibfnamefont
  {M.}~\bibnamefont {Toth}}, \ and\ \bibinfo {author} {\bibfnamefont
  {I.}~\bibnamefont {Aharonovich}},\ }\href {\doibase
  10.1021/acsphotonics.7b00977} {\bibfield  {journal} {\bibinfo  {journal} {ACS
  Photonics}\ }\textbf {\bibinfo {volume} {5}},\ \bibinfo {pages} {295}
  (\bibinfo {year} {2018})}\BibitemShut {NoStop}%
\bibitem [{\citenamefont {Exarhos}\ \emph {et~al.}(2017)\citenamefont
  {Exarhos}, \citenamefont {Hopper}, \citenamefont {Grote}, \citenamefont
  {Alkauskas},\ and\ \citenamefont {Bassett}}]{ExaHopGro17}%
  \BibitemOpen
  \bibfield  {author} {\bibinfo {author} {\bibfnamefont {A.~L.}\ \bibnamefont
  {Exarhos}}, \bibinfo {author} {\bibfnamefont {D.~A.}\ \bibnamefont {Hopper}},
  \bibinfo {author} {\bibfnamefont {R.~R.}\ \bibnamefont {Grote}}, \bibinfo
  {author} {\bibfnamefont {A.}~\bibnamefont {Alkauskas}}, \ and\ \bibinfo
  {author} {\bibfnamefont {L.~C.}\ \bibnamefont {Bassett}},\ }\href {\doibase
  10.1021/acsnano.7b00665} {\bibfield  {journal} {\bibinfo  {journal} {ACS
  Nano}\ }\textbf {\bibinfo {volume} {11}},\ \bibinfo {pages} {3328} (\bibinfo
  {year} {2017})}\BibitemShut {NoStop}%
\bibitem [{\citenamefont {Shotan}\ \emph {et~al.}(2016)\citenamefont {Shotan},
  \citenamefont {Jayakumar}, \citenamefont {Considine}, \citenamefont
  {Mackoit}, \citenamefont {Fedder}, \citenamefont {Wrachtrup}, \citenamefont
  {Alkauskas}, \citenamefont {Doherty}, \citenamefont {Menon},\ and\
  \citenamefont {Meriles}}]{ShoJayCon16}%
  \BibitemOpen
  \bibfield  {author} {\bibinfo {author} {\bibfnamefont {Z.}~\bibnamefont
  {Shotan}}, \bibinfo {author} {\bibfnamefont {H.}~\bibnamefont {Jayakumar}},
  \bibinfo {author} {\bibfnamefont {C.~R.}\ \bibnamefont {Considine}}, \bibinfo
  {author} {\bibfnamefont {M.}~\bibnamefont {Mackoit}}, \bibinfo {author}
  {\bibfnamefont {H.}~\bibnamefont {Fedder}}, \bibinfo {author} {\bibfnamefont
  {J.}~\bibnamefont {Wrachtrup}}, \bibinfo {author} {\bibfnamefont
  {A.}~\bibnamefont {Alkauskas}}, \bibinfo {author} {\bibfnamefont {M.~W.}\
  \bibnamefont {Doherty}}, \bibinfo {author} {\bibfnamefont {V.~M.}\
  \bibnamefont {Menon}}, \ and\ \bibinfo {author} {\bibfnamefont {C.~A.}\
  \bibnamefont {Meriles}},\ }\href {\doibase 10.1021/acsphotonics.6b00736}
  {\bibfield  {journal} {\bibinfo  {journal} {ACS Photonics}\ }\textbf
  {\bibinfo {volume} {3}},\ \bibinfo {pages} {2490} (\bibinfo {year}
  {2016})}\BibitemShut {NoStop}%
\bibitem [{\citenamefont {Tran}\ \emph {et~al.}(2016)\citenamefont {Tran},
  \citenamefont {Elbadawi}, \citenamefont {Totonjian}, \citenamefont {Lobo},
  \citenamefont {Grosso}, \citenamefont {Moon}, \citenamefont {Englund},
  \citenamefont {Ford}, \citenamefont {Aharonovich},\ and\ \citenamefont
  {Toth}}]{TraElbTot16}%
  \BibitemOpen
  \bibfield  {author} {\bibinfo {author} {\bibfnamefont {T.~T.}\ \bibnamefont
  {Tran}}, \bibinfo {author} {\bibfnamefont {C.}~\bibnamefont {Elbadawi}},
  \bibinfo {author} {\bibfnamefont {D.}~\bibnamefont {Totonjian}}, \bibinfo
  {author} {\bibfnamefont {C.~J.}\ \bibnamefont {Lobo}}, \bibinfo {author}
  {\bibfnamefont {G.}~\bibnamefont {Grosso}}, \bibinfo {author} {\bibfnamefont
  {H.}~\bibnamefont {Moon}}, \bibinfo {author} {\bibfnamefont {D.~R.}\
  \bibnamefont {Englund}}, \bibinfo {author} {\bibfnamefont {M.~J.}\
  \bibnamefont {Ford}}, \bibinfo {author} {\bibfnamefont {I.}~\bibnamefont
  {Aharonovich}}, \ and\ \bibinfo {author} {\bibfnamefont {M.}~\bibnamefont
  {Toth}},\ }\href {\doibase 10.1021/acsnano.6b03602} {\bibfield  {journal}
  {\bibinfo  {journal} {ACS Nano}\ }\textbf {\bibinfo {volume} {10}},\ \bibinfo
  {pages} {7331} (\bibinfo {year} {2016})}\BibitemShut {NoStop}%
\bibitem [{\citenamefont {Wigger}\ \emph {et~al.}(2019)\citenamefont {Wigger},
  \citenamefont {Schmidt}, \citenamefont {Pozo-Zamudio}, \citenamefont
  {Preu{ß}}, \citenamefont {Tonndorf}, \citenamefont {Schneider},
  \citenamefont {Steeger}, \citenamefont {Kern}, \citenamefont {Khodaei},
  \citenamefont {Sperling}, \citenamefont {de~Vasconcellos}, \citenamefont
  {Bratschitsch},\ and\ \citenamefont {Kuhn}}]{WigSchPoz19}%
  \BibitemOpen
  \bibfield  {author} {\bibinfo {author} {\bibfnamefont {D.}~\bibnamefont
  {Wigger}}, \bibinfo {author} {\bibfnamefont {R.}~\bibnamefont {Schmidt}},
  \bibinfo {author} {\bibfnamefont {O.~D.}\ \bibnamefont {Pozo-Zamudio}},
  \bibinfo {author} {\bibfnamefont {J.~A.}\ \bibnamefont {Preu{ß}}}, \bibinfo
  {author} {\bibfnamefont {P.}~\bibnamefont {Tonndorf}}, \bibinfo {author}
  {\bibfnamefont {R.}~\bibnamefont {Schneider}}, \bibinfo {author}
  {\bibfnamefont {P.}~\bibnamefont {Steeger}}, \bibinfo {author} {\bibfnamefont
  {J.}~\bibnamefont {Kern}}, \bibinfo {author} {\bibfnamefont {Y.}~\bibnamefont
  {Khodaei}}, \bibinfo {author} {\bibfnamefont {J.}~\bibnamefont {Sperling}},
  \bibinfo {author} {\bibfnamefont {S.~M.}\ \bibnamefont {de~Vasconcellos}},
  \bibinfo {author} {\bibfnamefont {R.}~\bibnamefont {Bratschitsch}}, \ and\
  \bibinfo {author} {\bibfnamefont {T.}~\bibnamefont {Kuhn}},\ }\href {\doibase
  10.1088/2053-1583/ab1188} {\bibfield  {journal} {\bibinfo  {journal} {2D
  Materials}\ }\textbf {\bibinfo {volume} {6}},\ \bibinfo {pages} {035006}
  (\bibinfo {year} {2019})}\BibitemShut {NoStop}%
\bibitem [{\citenamefont {Bommer}\ and\ \citenamefont
  {Becher}(2019)}]{BomBec19}%
  \BibitemOpen
  \bibfield  {author} {\bibinfo {author} {\bibfnamefont {A.}~\bibnamefont
  {Bommer}}\ and\ \bibinfo {author} {\bibfnamefont {C.}~\bibnamefont
  {Becher}},\ }\href {\doibase 10.1515/nanoph-2019-0123} {\bibfield  {journal}
  {\bibinfo  {journal} {Nanophotonics}\ }\textbf {\bibinfo {volume} {8}},\
  \bibinfo {pages} {2041} (\bibinfo {year} {2019})}\BibitemShut {NoStop}%
\bibitem [{\citenamefont {Dietrich}\ \emph {et~al.}(2020)\citenamefont
  {Dietrich}, \citenamefont {Doherty}, \citenamefont {Aharonovich},\ and\
  \citenamefont {Kubanek}}]{dietrich_solid_2019}%
  \BibitemOpen
  \bibfield  {author} {\bibinfo {author} {\bibfnamefont {A.}~\bibnamefont
  {Dietrich}}, \bibinfo {author} {\bibfnamefont {M.~W.}\ \bibnamefont
  {Doherty}}, \bibinfo {author} {\bibfnamefont {I.}~\bibnamefont
  {Aharonovich}}, \ and\ \bibinfo {author} {\bibfnamefont {A.}~\bibnamefont
  {Kubanek}},\ }\href {\doibase 10.1103/PhysRevB.101.081401} {\bibfield
  {journal} {\bibinfo  {journal} {Phys. Rev. B}\ }\textbf {\bibinfo {volume}
  {101}},\ \bibinfo {pages} {081401(R)} (\bibinfo {year} {2020})}\BibitemShut
  {NoStop}%
\bibitem [{\citenamefont {Atat\"ure}\ \emph {et~al.}(2018)\citenamefont
  {Atat\"ure}, \citenamefont {Englund}, \citenamefont {Vamivakas},
  \citenamefont {Lee},\ and\ \citenamefont
  {Wrachtrup}}]{atature_material_2018}%
  \BibitemOpen
  \bibfield  {author} {\bibinfo {author} {\bibfnamefont {M.}~\bibnamefont
  {Atat\"ure}}, \bibinfo {author} {\bibfnamefont {D.}~\bibnamefont {Englund}},
  \bibinfo {author} {\bibfnamefont {N.}~\bibnamefont {Vamivakas}}, \bibinfo
  {author} {\bibfnamefont {S.-Y.}\ \bibnamefont {Lee}}, \ and\ \bibinfo
  {author} {\bibfnamefont {J.}~\bibnamefont {Wrachtrup}},\ }\href {\doibase
  10.1038/s41578-018-0008-9} {\bibfield  {journal} {\bibinfo  {journal} {Nature
  Reviews Materials}\ }\textbf {\bibinfo {volume} {3}},\ \bibinfo {pages} {38}
  (\bibinfo {year} {2018})}\BibitemShut {NoStop}%
\bibitem [{\citenamefont {Abdi}\ \emph {et~al.}(2017)\citenamefont {Abdi},
  \citenamefont {Hwang}, \citenamefont {Aghtar},\ and\ \citenamefont
  {Plenio}}]{abdi_spin-mechanical_2017}%
  \BibitemOpen
  \bibfield  {author} {\bibinfo {author} {\bibfnamefont {M.}~\bibnamefont
  {Abdi}}, \bibinfo {author} {\bibfnamefont {M.-J.}\ \bibnamefont {Hwang}},
  \bibinfo {author} {\bibfnamefont {M.}~\bibnamefont {Aghtar}}, \ and\ \bibinfo
  {author} {\bibfnamefont {M.~B.}\ \bibnamefont {Plenio}},\ }\href {\doibase
  10.1103/PhysRevLett.119.233602} {\bibfield  {journal} {\bibinfo  {journal}
  {Physical Review Letters}\ }\textbf {\bibinfo {volume} {119}},\ \bibinfo
  {pages} {233602} (\bibinfo {year} {2017})}\BibitemShut {NoStop}%
\bibitem [{\citenamefont {Abdi}\ and\ \citenamefont
  {Plenio}(2019)}]{abdi_quantum_2019}%
  \BibitemOpen
  \bibfield  {author} {\bibinfo {author} {\bibfnamefont {M.}~\bibnamefont
  {Abdi}}\ and\ \bibinfo {author} {\bibfnamefont {M.~B.}\ \bibnamefont
  {Plenio}},\ }\href {\doibase 10.1103/PhysRevLett.122.023602} {\bibfield
  {journal} {\bibinfo  {journal} {Physical Review Letters}\ }\textbf {\bibinfo
  {volume} {122}},\ \bibinfo {pages} {023602} (\bibinfo {year}
  {2019})}\BibitemShut {NoStop}%
\bibitem [{\citenamefont {Xia}\ \emph {et~al.}(2019)\citenamefont {Xia},
  \citenamefont {Li}, \citenamefont {Kim}, \citenamefont {Bao}, \citenamefont
  {Gong}, \citenamefont {Yang}, \citenamefont {Wang},\ and\ \citenamefont
  {Zhang}}]{XiaLiKim19}%
  \BibitemOpen
  \bibfield  {author} {\bibinfo {author} {\bibfnamefont {Y.}~\bibnamefont
  {Xia}}, \bibinfo {author} {\bibfnamefont {Q.}~\bibnamefont {Li}}, \bibinfo
  {author} {\bibfnamefont {J.}~\bibnamefont {Kim}}, \bibinfo {author}
  {\bibfnamefont {W.}~\bibnamefont {Bao}}, \bibinfo {author} {\bibfnamefont
  {C.}~\bibnamefont {Gong}}, \bibinfo {author} {\bibfnamefont {S.}~\bibnamefont
  {Yang}}, \bibinfo {author} {\bibfnamefont {Y.}~\bibnamefont {Wang}}, \ and\
  \bibinfo {author} {\bibfnamefont {X.}~\bibnamefont {Zhang}},\ }\href
  {\doibase 10.1021/acs.nanolett.9b02640} {\bibfield  {journal} {\bibinfo
  {journal} {Nano Letters}\ }\textbf {\bibinfo {volume} {19}},\ \bibinfo
  {pages} {7100} (\bibinfo {year} {2019})}\BibitemShut {NoStop}%
\bibitem [{\citenamefont {Nikolay}\ \emph {et~al.}(2019)\citenamefont
  {Nikolay}, \citenamefont {Mendelson}, \citenamefont {\"Ozelci}, \citenamefont
  {Sontheimer}, \citenamefont {B\"ohm}, \citenamefont {Kewes}, \citenamefont
  {Toth}, \citenamefont {Aharonovich},\ and\ \citenamefont
  {Benson}}]{NikMenOze19}%
  \BibitemOpen
  \bibfield  {author} {\bibinfo {author} {\bibfnamefont {N.}~\bibnamefont
  {Nikolay}}, \bibinfo {author} {\bibfnamefont {N.}~\bibnamefont {Mendelson}},
  \bibinfo {author} {\bibfnamefont {E.}~\bibnamefont {\"Ozelci}}, \bibinfo
  {author} {\bibfnamefont {B.}~\bibnamefont {Sontheimer}}, \bibinfo {author}
  {\bibfnamefont {F.}~\bibnamefont {B\"ohm}}, \bibinfo {author} {\bibfnamefont
  {G.}~\bibnamefont {Kewes}}, \bibinfo {author} {\bibfnamefont
  {M.}~\bibnamefont {Toth}}, \bibinfo {author} {\bibfnamefont {I.}~\bibnamefont
  {Aharonovich}}, \ and\ \bibinfo {author} {\bibfnamefont {O.}~\bibnamefont
  {Benson}},\ }\href {\doibase 10.1364/OPTICA.6.001084} {\bibfield  {journal}
  {\bibinfo  {journal} {Optica}\ }\textbf {\bibinfo {volume} {6}},\ \bibinfo
  {pages} {1084} (\bibinfo {year} {2019})}\BibitemShut {NoStop}%
\bibitem [{\citenamefont {Museur}\ \emph {et~al.}(2008)\citenamefont {Museur},
  \citenamefont {Feldbach},\ and\ \citenamefont {Kanaev}}]{MusFelKan08}%
  \BibitemOpen
  \bibfield  {author} {\bibinfo {author} {\bibfnamefont {L.}~\bibnamefont
  {Museur}}, \bibinfo {author} {\bibfnamefont {E.}~\bibnamefont {Feldbach}}, \
  and\ \bibinfo {author} {\bibfnamefont {A.}~\bibnamefont {Kanaev}},\ }\href
  {\doibase 10.1103/PhysRevB.78.155204} {\bibfield  {journal} {\bibinfo
  {journal} {Physical Review B}\ }\textbf {\bibinfo {volume} {78}},\ \bibinfo
  {pages} {155204} (\bibinfo {year} {2008})}\BibitemShut {NoStop}%
\bibitem [{\citenamefont {Bourrellier}\ \emph {et~al.}(2016)\citenamefont
  {Bourrellier}, \citenamefont {Meuret}, \citenamefont {Tararan}, \citenamefont
  {St\'{e}phan}, \citenamefont {Kociak}, \citenamefont {Tizei},\ and\
  \citenamefont {Zobelli}}]{BouMeuTar16}%
  \BibitemOpen
  \bibfield  {author} {\bibinfo {author} {\bibfnamefont {R.}~\bibnamefont
  {Bourrellier}}, \bibinfo {author} {\bibfnamefont {S.}~\bibnamefont {Meuret}},
  \bibinfo {author} {\bibfnamefont {A.}~\bibnamefont {Tararan}}, \bibinfo
  {author} {\bibfnamefont {O.}~\bibnamefont {St\'{e}phan}}, \bibinfo {author}
  {\bibfnamefont {M.}~\bibnamefont {Kociak}}, \bibinfo {author} {\bibfnamefont
  {L.~H.~G.}\ \bibnamefont {Tizei}}, \ and\ \bibinfo {author} {\bibfnamefont
  {A.}~\bibnamefont {Zobelli}},\ }\href {\doibase 10.1021/acs.nanolett.6b01368}
  {\bibfield  {journal} {\bibinfo  {journal} {Nano Letters}\ }\textbf {\bibinfo
  {volume} {16}},\ \bibinfo {pages} {4317} (\bibinfo {year}
  {2016})}\BibitemShut {NoStop}%
\bibitem [{\citenamefont {Tsushima}\ \emph {et~al.}(2018)\citenamefont
  {Tsushima}, \citenamefont {Tsujimura},\ and\ \citenamefont
  {Uchino}}]{TsuTsuUch18}%
  \BibitemOpen
  \bibfield  {author} {\bibinfo {author} {\bibfnamefont {E.}~\bibnamefont
  {Tsushima}}, \bibinfo {author} {\bibfnamefont {T.}~\bibnamefont {Tsujimura}},
  \ and\ \bibinfo {author} {\bibfnamefont {T.}~\bibnamefont {Uchino}},\ }\href
  {\doibase 10.1063/1.5038168} {\bibfield  {journal} {\bibinfo  {journal}
  {Applied Physics Letters}\ }\textbf {\bibinfo {volume} {113}},\ \bibinfo
  {pages} {031903} (\bibinfo {year} {2018})}\BibitemShut {NoStop}%
\bibitem [{\citenamefont {Vuong}\ \emph {et~al.}(2016)\citenamefont {Vuong},
  \citenamefont {Cassabois}, \citenamefont {Valvin}, \citenamefont {Ouerghi},
  \citenamefont {Chassagneux}, \citenamefont {Voisin},\ and\ \citenamefont
  {Gil}}]{VuoCasVal16}%
  \BibitemOpen
  \bibfield  {author} {\bibinfo {author} {\bibfnamefont {T.~Q.~P.}\
  \bibnamefont {Vuong}}, \bibinfo {author} {\bibfnamefont {G.}~\bibnamefont
  {Cassabois}}, \bibinfo {author} {\bibfnamefont {P.}~\bibnamefont {Valvin}},
  \bibinfo {author} {\bibfnamefont {A.}~\bibnamefont {Ouerghi}}, \bibinfo
  {author} {\bibfnamefont {Y.}~\bibnamefont {Chassagneux}}, \bibinfo {author}
  {\bibfnamefont {C.}~\bibnamefont {Voisin}}, \ and\ \bibinfo {author}
  {\bibfnamefont {B.}~\bibnamefont {Gil}},\ }\href {\doibase
  10.1103/PhysRevLett.117.097402} {\bibfield  {journal} {\bibinfo  {journal}
  {Physical Review Letters}\ }\textbf {\bibinfo {volume} {117}},\ \bibinfo
  {pages} {097402} (\bibinfo {year} {2016})}\BibitemShut {NoStop}%
\bibitem [{\citenamefont {Tawfik}\ \emph {et~al.}(2017)\citenamefont {Tawfik},
  \citenamefont {Ali}, \citenamefont {Fronzi}, \citenamefont {Kianinia},
  \citenamefont {Tran}, \citenamefont {Stampfl}, \citenamefont {Aharonovich},
  \citenamefont {Toth},\ and\ \citenamefont {Ford}}]{TawAliFro17}%
  \BibitemOpen
  \bibfield  {author} {\bibinfo {author} {\bibfnamefont {S.~A.}\ \bibnamefont
  {Tawfik}}, \bibinfo {author} {\bibfnamefont {S.}~\bibnamefont {Ali}},
  \bibinfo {author} {\bibfnamefont {M.}~\bibnamefont {Fronzi}}, \bibinfo
  {author} {\bibfnamefont {M.}~\bibnamefont {Kianinia}}, \bibinfo {author}
  {\bibfnamefont {T.~T.}\ \bibnamefont {Tran}}, \bibinfo {author}
  {\bibfnamefont {C.}~\bibnamefont {Stampfl}}, \bibinfo {author} {\bibfnamefont
  {I.}~\bibnamefont {Aharonovich}}, \bibinfo {author} {\bibfnamefont
  {M.}~\bibnamefont {Toth}}, \ and\ \bibinfo {author} {\bibfnamefont {M.~J.}\
  \bibnamefont {Ford}},\ }\href {\doibase 10.1039/C7NR04270A} {\bibfield
  {journal} {\bibinfo  {journal} {Nanoscale}\ }\textbf {\bibinfo {volume}
  {9}},\ \bibinfo {pages} {13575} (\bibinfo {year} {2017})}\BibitemShut
  {NoStop}%
\bibitem [{\citenamefont {Weston}\ \emph {et~al.}(2018)\citenamefont {Weston},
  \citenamefont {Wickramaratne}, \citenamefont {Mackoit}, \citenamefont
  {Alkauskas},\ and\ \citenamefont {Van~de Walle}}]{WesWicMac18}%
  \BibitemOpen
  \bibfield  {author} {\bibinfo {author} {\bibfnamefont {L.}~\bibnamefont
  {Weston}}, \bibinfo {author} {\bibfnamefont {D.}~\bibnamefont
  {Wickramaratne}}, \bibinfo {author} {\bibfnamefont {M.}~\bibnamefont
  {Mackoit}}, \bibinfo {author} {\bibfnamefont {A.}~\bibnamefont {Alkauskas}},
  \ and\ \bibinfo {author} {\bibfnamefont {C.~G.}\ \bibnamefont {Van~de
  Walle}},\ }\href {\doibase 10.1103/PhysRevB.97.214104} {\bibfield  {journal}
  {\bibinfo  {journal} {Physical Review B}\ }\textbf {\bibinfo {volume} {97}},\
  \bibinfo {pages} {214104} (\bibinfo {year} {2018})}\BibitemShut {NoStop}%
\bibitem [{\citenamefont {Feldman}\ \emph {et~al.}(2019)\citenamefont
  {Feldman}, \citenamefont {Puretzky}, \citenamefont {Lindsay}, \citenamefont
  {Tucker}, \citenamefont {Briggs}, \citenamefont {Evans}, \citenamefont
  {Haglund},\ and\ \citenamefont {Lawrie}}]{FelPurLin19}%
  \BibitemOpen
  \bibfield  {author} {\bibinfo {author} {\bibfnamefont {M.~A.}\ \bibnamefont
  {Feldman}}, \bibinfo {author} {\bibfnamefont {A.}~\bibnamefont {Puretzky}},
  \bibinfo {author} {\bibfnamefont {L.}~\bibnamefont {Lindsay}}, \bibinfo
  {author} {\bibfnamefont {E.}~\bibnamefont {Tucker}}, \bibinfo {author}
  {\bibfnamefont {D.~P.}\ \bibnamefont {Briggs}}, \bibinfo {author}
  {\bibfnamefont {P.~G.}\ \bibnamefont {Evans}}, \bibinfo {author}
  {\bibfnamefont {R.~F.}\ \bibnamefont {Haglund}}, \ and\ \bibinfo {author}
  {\bibfnamefont {B.~J.}\ \bibnamefont {Lawrie}},\ }\href {\doibase
  10.1103/PhysRevB.99.020101} {\bibfield  {journal} {\bibinfo  {journal}
  {Physical Review B}\ }\textbf {\bibinfo {volume} {99}},\ \bibinfo {pages}
  {020101(R)} (\bibinfo {year} {2019})}\BibitemShut {NoStop}%
\bibitem [{\citenamefont {Grosso}\ \emph {et~al.}(2020)\citenamefont {Grosso},
  \citenamefont {Moon}, \citenamefont {Ciccarino}, \citenamefont {Flick},
  \citenamefont {Mendelson}, \citenamefont {Mennel}, \citenamefont {Toth},
  \citenamefont {Aharonovich}, \citenamefont {Narang},\ and\ \citenamefont
  {Englund}}]{GroMooCic20}%
  \BibitemOpen
  \bibfield  {author} {\bibinfo {author} {\bibfnamefont {G.}~\bibnamefont
  {Grosso}}, \bibinfo {author} {\bibfnamefont {H.}~\bibnamefont {Moon}},
  \bibinfo {author} {\bibfnamefont {C.~J.}\ \bibnamefont {Ciccarino}}, \bibinfo
  {author} {\bibfnamefont {J.}~\bibnamefont {Flick}}, \bibinfo {author}
  {\bibfnamefont {N.}~\bibnamefont {Mendelson}}, \bibinfo {author}
  {\bibfnamefont {L.}~\bibnamefont {Mennel}}, \bibinfo {author} {\bibfnamefont
  {M.}~\bibnamefont {Toth}}, \bibinfo {author} {\bibfnamefont {I.}~\bibnamefont
  {Aharonovich}}, \bibinfo {author} {\bibfnamefont {P.}~\bibnamefont {Narang}},
  \ and\ \bibinfo {author} {\bibfnamefont {D.~R.}\ \bibnamefont {Englund}},\
  }\href {\doibase 10.1021/acsphotonics.9b01789} {\bibfield  {journal}
  {\bibinfo  {journal} {ACS Photonics}\ }\textbf {\bibinfo {volume} {7}},\
  \bibinfo {pages} {1410} (\bibinfo {year} {2020})}\BibitemShut {NoStop}%
\bibitem [{\citenamefont {Markham}(1959)}]{Mar59}%
  \BibitemOpen
  \bibfield  {author} {\bibinfo {author} {\bibfnamefont {J.~J.}\ \bibnamefont
  {Markham}},\ }\href {\doibase 10.1103/RevModPhys.31.956} {\bibfield
  {journal} {\bibinfo  {journal} {Review Modern Physics}\ }\textbf {\bibinfo
  {volume} {31}},\ \bibinfo {pages} {956} (\bibinfo {year} {1959})}\BibitemShut
  {NoStop}%
\bibitem [{Note1()}]{Note1}%
  \BibitemOpen
  \bibinfo {note} {We adhere to the convention that the cohesive energy is
  defined as the energy \protect \emph {gained} upon formation from the atomic
  states and hence commonly a \protect \emph {positive} quantity.}\BibitemShut
  {Stop}%
\bibitem [{\citenamefont {Lax}(1952)}]{Lax52}%
  \BibitemOpen
  \bibfield  {author} {\bibinfo {author} {\bibfnamefont {M.}~\bibnamefont
  {Lax}},\ }\href {\doibase 10.1063/1.1700283} {\bibfield  {journal} {\bibinfo
  {journal} {The Journal of Chemical Physics}\ }\textbf {\bibinfo {volume}
  {20}},\ \bibinfo {pages} {1752} (\bibinfo {year} {1952})}\BibitemShut
  {NoStop}%
\bibitem [{\citenamefont {Kubo}\ and\ \citenamefont
  {Toyozawa}(1955)}]{KubToy55}%
  \BibitemOpen
  \bibfield  {author} {\bibinfo {author} {\bibfnamefont {R.}~\bibnamefont
  {Kubo}}\ and\ \bibinfo {author} {\bibfnamefont {Y.}~\bibnamefont
  {Toyozawa}},\ }\href {\doibase 10.1143/PTP.13.160} {\bibfield  {journal}
  {\bibinfo  {journal} {Progress of Theoretical Physics}\ }\textbf {\bibinfo
  {volume} {13}},\ \bibinfo {pages} {160} (\bibinfo {year} {1955})}\BibitemShut
  {NoStop}%
\bibitem [{\citenamefont {Alkauskas}\ \emph {et~al.}(2014)\citenamefont
  {Alkauskas}, \citenamefont {Buckley}, \citenamefont {Awschalom},\ and\
  \citenamefont {de~Walle}}]{AlkBucAws14}%
  \BibitemOpen
  \bibfield  {author} {\bibinfo {author} {\bibfnamefont {A.}~\bibnamefont
  {Alkauskas}}, \bibinfo {author} {\bibfnamefont {B.~B.}\ \bibnamefont
  {Buckley}}, \bibinfo {author} {\bibfnamefont {D.~D.}\ \bibnamefont
  {Awschalom}}, \ and\ \bibinfo {author} {\bibfnamefont {C.~G.~V.}\
  \bibnamefont {de~Walle}},\ }\href {\doibase 10.1088/1367-2630/16/7/073026}
  {\bibfield  {journal} {\bibinfo  {journal} {New Journal of Physics}\ }\textbf
  {\bibinfo {volume} {16}},\ \bibinfo {pages} {073026} (\bibinfo {year}
  {2014})}\BibitemShut {NoStop}%
\bibitem [{\citenamefont {Kresse}\ and\ \citenamefont
  {Furthm{\"u}ller}(1996)}]{KreFur96}%
  \BibitemOpen
  \bibfield  {author} {\bibinfo {author} {\bibfnamefont {G.}~\bibnamefont
  {Kresse}}\ and\ \bibinfo {author} {\bibfnamefont {J.}~\bibnamefont
  {Furthm{\"u}ller}},\ }\href {\doibase 10.1103/PhysRevB.54.11169} {\bibfield
  {journal} {\bibinfo  {journal} {Physical Review B}\ }\textbf {\bibinfo
  {volume} {54}},\ \bibinfo {pages} {11169} (\bibinfo {year}
  {1996})}\BibitemShut {NoStop}%
\bibitem [{\citenamefont {Bl\"ochl}(1994)}]{Blo94}%
  \BibitemOpen
  \bibfield  {author} {\bibinfo {author} {\bibfnamefont {P.~E.}\ \bibnamefont
  {Bl\"ochl}},\ }\href {\doibase 10.1103/PhysRevB.50.17953} {\bibfield
  {journal} {\bibinfo  {journal} {Physical Review B}\ }\textbf {\bibinfo
  {volume} {50}},\ \bibinfo {pages} {17953} (\bibinfo {year}
  {1994})}\BibitemShut {NoStop}%
\bibitem [{\citenamefont {Kresse}\ and\ \citenamefont
  {Joubert}(1999)}]{KreJou99}%
  \BibitemOpen
  \bibfield  {author} {\bibinfo {author} {\bibfnamefont {G.}~\bibnamefont
  {Kresse}}\ and\ \bibinfo {author} {\bibfnamefont {D.}~\bibnamefont
  {Joubert}},\ }\href {\doibase 10.1103/PhysRevB.59.1758} {\bibfield  {journal}
  {\bibinfo  {journal} {Physical Review B}\ }\textbf {\bibinfo {volume} {59}},\
  \bibinfo {pages} {1758} (\bibinfo {year} {1999})}\BibitemShut {NoStop}%
\bibitem [{\citenamefont {Kresse}\ and\ \citenamefont
  {Hafner}(1993)}]{KreHaf93}%
  \BibitemOpen
  \bibfield  {author} {\bibinfo {author} {\bibfnamefont {G.}~\bibnamefont
  {Kresse}}\ and\ \bibinfo {author} {\bibfnamefont {J.}~\bibnamefont
  {Hafner}},\ }\href {\doibase 10.1103/PhysRevB.47.558} {\bibfield  {journal}
  {\bibinfo  {journal} {Physical Review B}\ }\textbf {\bibinfo {volume} {47}},\
  \bibinfo {pages} {558} (\bibinfo {year} {1993})}\BibitemShut {NoStop}%
\bibitem [{\citenamefont {Perdew}\ \emph {et~al.}(1996)\citenamefont {Perdew},
  \citenamefont {Burke},\ and\ \citenamefont {Ernzerhof}}]{PerBurErn96}%
  \BibitemOpen
  \bibfield  {author} {\bibinfo {author} {\bibfnamefont {J.~P.}\ \bibnamefont
  {Perdew}}, \bibinfo {author} {\bibfnamefont {K.}~\bibnamefont {Burke}}, \
  and\ \bibinfo {author} {\bibfnamefont {M.}~\bibnamefont {Ernzerhof}},\ }\href
  {\doibase 10.1103/PhysRevLett.77.3865} {\bibfield  {journal} {\bibinfo
  {journal} {Physical Review Letters}\ }\textbf {\bibinfo {volume} {77}},\
  \bibinfo {pages} {3865} (\bibinfo {year} {1996})}\BibitemShut {NoStop}%
\bibitem [{\citenamefont {Heyd}\ \emph {et~al.}(2003)\citenamefont {Heyd},
  \citenamefont {Scuseria},\ and\ \citenamefont {Ernzerhof}}]{HeyScuErn03}%
  \BibitemOpen
  \bibfield  {author} {\bibinfo {author} {\bibfnamefont {J.}~\bibnamefont
  {Heyd}}, \bibinfo {author} {\bibfnamefont {G.~E.}\ \bibnamefont {Scuseria}},
  \ and\ \bibinfo {author} {\bibfnamefont {M.}~\bibnamefont {Ernzerhof}},\
  }\href {\doibase 10.1063/1.1564060} {\bibfield  {journal} {\bibinfo
  {journal} {The Journal of Chemical Physics}\ }\textbf {\bibinfo {volume}
  {118}},\ \bibinfo {pages} {8207} (\bibinfo {year} {2003})}\BibitemShut
  {NoStop}%
\bibitem [{\citenamefont {Komsa}\ \emph {et~al.}(2014)\citenamefont {Komsa},
  \citenamefont {Berseneva}, \citenamefont {Krasheninnikov},\ and\
  \citenamefont {Nieminen}}]{KomBerArk14}%
  \BibitemOpen
  \bibfield  {author} {\bibinfo {author} {\bibfnamefont {H.-P.}\ \bibnamefont
  {Komsa}}, \bibinfo {author} {\bibfnamefont {N.}~\bibnamefont {Berseneva}},
  \bibinfo {author} {\bibfnamefont {A.~V.}\ \bibnamefont {Krasheninnikov}}, \
  and\ \bibinfo {author} {\bibfnamefont {R.~M.}\ \bibnamefont {Nieminen}},\
  }\href {\doibase 10.1103/PhysRevX.4.031044} {\bibfield  {journal} {\bibinfo
  {journal} {Physical Review X}\ }\textbf {\bibinfo {volume} {4}},\ \bibinfo
  {pages} {031044} (\bibinfo {year} {2014})}\BibitemShut {NoStop}%
\bibitem [{\citenamefont {Komsa}(2020)}]{Komsa2020}%
  \BibitemOpen
  \bibfield  {author} {\bibinfo {author} {\bibfnamefont {H.-P.}\ \bibnamefont
  {Komsa}},\ }\href@noop {} {} (\bibinfo {year} {2020}),\ \bibinfo {note}
  {private communication}\BibitemShut {NoStop}%
\bibitem [{\citenamefont {Togo}\ and\ \citenamefont {Tanaka}(2015)}]{TogTan15}%
  \BibitemOpen
  \bibfield  {author} {\bibinfo {author} {\bibfnamefont {A.}~\bibnamefont
  {Togo}}\ and\ \bibinfo {author} {\bibfnamefont {I.}~\bibnamefont {Tanaka}},\
  }\href@noop {} {\bibfield  {journal} {\bibinfo  {journal} {Scripta
  Materialia}\ }\textbf {\bibinfo {volume} {108}},\ \bibinfo {pages} {1}
  (\bibinfo {year} {2015})}\BibitemShut {NoStop}%
\bibitem [{\citenamefont {Paszkowicz}\ \emph {et~al.}(2002)\citenamefont
  {Paszkowicz}, \citenamefont {Pelka}, \citenamefont {Knapp}, \citenamefont
  {Szyszko},\ and\ \citenamefont {Podsiadlo}}]{PasPelKna02}%
  \BibitemOpen
  \bibfield  {author} {\bibinfo {author} {\bibfnamefont {W.}~\bibnamefont
  {Paszkowicz}}, \bibinfo {author} {\bibfnamefont {J.}~\bibnamefont {Pelka}},
  \bibinfo {author} {\bibfnamefont {M.}~\bibnamefont {Knapp}}, \bibinfo
  {author} {\bibfnamefont {T.}~\bibnamefont {Szyszko}}, \ and\ \bibinfo
  {author} {\bibfnamefont {S.}~\bibnamefont {Podsiadlo}},\ }\href {\doibase
  10.1007/s003390100999} {\bibfield  {journal} {\bibinfo  {journal} {Applied
  Physics A}\ }\textbf {\bibinfo {volume} {75}},\ \bibinfo {pages} {431}
  (\bibinfo {year} {2002})}\BibitemShut {NoStop}%
\bibitem [{\citenamefont {Haastrup}\ \emph {et~al.}(2018)\citenamefont
  {Haastrup}, \citenamefont {Strange}, \citenamefont {Pandey}, \citenamefont
  {Deilmann}, \citenamefont {Schmidt}, \citenamefont {Hinsche}, \citenamefont
  {Gjerding}, \citenamefont {Torelli}, \citenamefont {Larsen}, \citenamefont
  {Riis-Jensen}, \citenamefont {Gath}, \citenamefont {Jacobsen}, \citenamefont
  {Mortensen}, \citenamefont {Olsen},\ and\ \citenamefont
  {Thygesen}}]{HaaStrPan18}%
  \BibitemOpen
  \bibfield  {author} {\bibinfo {author} {\bibfnamefont {S.}~\bibnamefont
  {Haastrup}}, \bibinfo {author} {\bibfnamefont {M.}~\bibnamefont {Strange}},
  \bibinfo {author} {\bibfnamefont {M.}~\bibnamefont {Pandey}}, \bibinfo
  {author} {\bibfnamefont {T.}~\bibnamefont {Deilmann}}, \bibinfo {author}
  {\bibfnamefont {P.~S.}\ \bibnamefont {Schmidt}}, \bibinfo {author}
  {\bibfnamefont {N.~F.}\ \bibnamefont {Hinsche}}, \bibinfo {author}
  {\bibfnamefont {M.~N.}\ \bibnamefont {Gjerding}}, \bibinfo {author}
  {\bibfnamefont {D.}~\bibnamefont {Torelli}}, \bibinfo {author} {\bibfnamefont
  {P.~M.}\ \bibnamefont {Larsen}}, \bibinfo {author} {\bibfnamefont {A.~C.}\
  \bibnamefont {Riis-Jensen}}, \bibinfo {author} {\bibfnamefont
  {J.}~\bibnamefont {Gath}}, \bibinfo {author} {\bibfnamefont {K.~W.}\
  \bibnamefont {Jacobsen}}, \bibinfo {author} {\bibfnamefont {J.~J.}\
  \bibnamefont {Mortensen}}, \bibinfo {author} {\bibfnamefont {T.}~\bibnamefont
  {Olsen}}, \ and\ \bibinfo {author} {\bibfnamefont {K.~S.}\ \bibnamefont
  {Thygesen}},\ }\href {\doibase 10.1088/2053-1583/aacfc1} {\bibfield
  {journal} {\bibinfo  {journal} {2D Materials}\ }\textbf {\bibinfo {volume}
  {5}},\ \bibinfo {pages} {042002} (\bibinfo {year} {2018})}\BibitemShut
  {NoStop}%
\bibitem [{\citenamefont {Elias}\ \emph {et~al.}(2019)\citenamefont {Elias},
  \citenamefont {Valvin}, \citenamefont {Pelini}, \citenamefont {Summerfield},
  \citenamefont {Mellor}, \citenamefont {Cheng}, \citenamefont {Eaves},
  \citenamefont {Foxon}, \citenamefont {Beton}, \citenamefont {Novikov},
  \citenamefont {Gil},\ and\ \citenamefont {Cassabois}}]{ElaValPel19}%
  \BibitemOpen
  \bibfield  {author} {\bibinfo {author} {\bibfnamefont {C.}~\bibnamefont
  {Elias}}, \bibinfo {author} {\bibfnamefont {P.}~\bibnamefont {Valvin}},
  \bibinfo {author} {\bibfnamefont {T.}~\bibnamefont {Pelini}}, \bibinfo
  {author} {\bibfnamefont {A.}~\bibnamefont {Summerfield}}, \bibinfo {author}
  {\bibfnamefont {C.~J.}\ \bibnamefont {Mellor}}, \bibinfo {author}
  {\bibfnamefont {T.~S.}\ \bibnamefont {Cheng}}, \bibinfo {author}
  {\bibfnamefont {L.}~\bibnamefont {Eaves}}, \bibinfo {author} {\bibfnamefont
  {C.~T.}\ \bibnamefont {Foxon}}, \bibinfo {author} {\bibfnamefont {P.~H.}\
  \bibnamefont {Beton}}, \bibinfo {author} {\bibfnamefont {S.~V.}\ \bibnamefont
  {Novikov}}, \bibinfo {author} {\bibfnamefont {B.}~\bibnamefont {Gil}}, \ and\
  \bibinfo {author} {\bibfnamefont {G.}~\bibnamefont {Cassabois}},\ }\href
  {\doibase 10.1038/s41467-019-10610-5} {\bibfield  {journal} {\bibinfo
  {journal} {Nature Communications}\ }\textbf {\bibinfo {volume} {10}},\
  \bibinfo {pages} {2639} (\bibinfo {year} {2019})}\BibitemShut {NoStop}%
\bibitem [{\citenamefont {Cassabois}\ \emph {et~al.}(2016)\citenamefont
  {Cassabois}, \citenamefont {Valvin},\ and\ \citenamefont
  {Gil}}]{CasValGil16}%
  \BibitemOpen
  \bibfield  {author} {\bibinfo {author} {\bibfnamefont {G.}~\bibnamefont
  {Cassabois}}, \bibinfo {author} {\bibfnamefont {P.}~\bibnamefont {Valvin}}, \
  and\ \bibinfo {author} {\bibfnamefont {B.}~\bibnamefont {Gil}},\ }\href
  {\doibase 10.1038/nphoton.2015.277} {\bibfield  {journal} {\bibinfo
  {journal} {Nature Photonics}\ }\textbf {\bibinfo {volume} {10}},\ \bibinfo
  {pages} {262} (\bibinfo {year} {2016})}\BibitemShut {NoStop}%
\bibitem [{\citenamefont {Arnaud}\ \emph {et~al.}(2006)\citenamefont {Arnaud},
  \citenamefont {Leb\`egue}, \citenamefont {Rabiller},\ and\ \citenamefont
  {Alouani}}]{ArnLebRab06}%
  \BibitemOpen
  \bibfield  {author} {\bibinfo {author} {\bibfnamefont {B.}~\bibnamefont
  {Arnaud}}, \bibinfo {author} {\bibfnamefont {S.}~\bibnamefont {Leb\`egue}},
  \bibinfo {author} {\bibfnamefont {P.}~\bibnamefont {Rabiller}}, \ and\
  \bibinfo {author} {\bibfnamefont {M.}~\bibnamefont {Alouani}},\ }\href
  {\doibase 10.1103/PhysRevLett.96.026402} {\bibfield  {journal} {\bibinfo
  {journal} {Physical Review Letters}\ }\textbf {\bibinfo {volume} {96}},\
  \bibinfo {pages} {026402} (\bibinfo {year} {2006})}\BibitemShut {NoStop}%
\bibitem [{\citenamefont {Tutchton}\ \emph {et~al.}(2018)\citenamefont
  {Tutchton}, \citenamefont {Marchbanks},\ and\ \citenamefont
  {Wu}}]{TutMarWu18}%
  \BibitemOpen
  \bibfield  {author} {\bibinfo {author} {\bibfnamefont {R.}~\bibnamefont
  {Tutchton}}, \bibinfo {author} {\bibfnamefont {C.}~\bibnamefont
  {Marchbanks}}, \ and\ \bibinfo {author} {\bibfnamefont {Z.}~\bibnamefont
  {Wu}},\ }\href {\doibase 10.1103/PhysRevB.97.205104} {\bibfield  {journal}
  {\bibinfo  {journal} {Phys. Rev. B}\ }\textbf {\bibinfo {volume} {97}},\
  \bibinfo {pages} {205104} (\bibinfo {year} {2018})}\BibitemShut {NoStop}%
\bibitem [{Note2()}]{Note2}%
  \BibitemOpen
  \bibinfo {note} {Since \gls {lo-to} splitting is absent in 2D materials \cite
  {SohGibCal17}, the non-analytical contribution to the force constant matrix
  has been omitted.}\BibitemShut {Stop}%
\bibitem [{\citenamefont {Carrete}\ \emph {et~al.}(2016)\citenamefont
  {Carrete}, \citenamefont {Li}, \citenamefont {Lindsay}, \citenamefont
  {Broido}, \citenamefont {Gallego},\ and\ \citenamefont {Mingo}}]{CarLiLin16}%
  \BibitemOpen
  \bibfield  {author} {\bibinfo {author} {\bibfnamefont {J.}~\bibnamefont
  {Carrete}}, \bibinfo {author} {\bibfnamefont {W.}~\bibnamefont {Li}},
  \bibinfo {author} {\bibfnamefont {L.}~\bibnamefont {Lindsay}}, \bibinfo
  {author} {\bibfnamefont {D.~A.}\ \bibnamefont {Broido}}, \bibinfo {author}
  {\bibfnamefont {L.~J.}\ \bibnamefont {Gallego}}, \ and\ \bibinfo {author}
  {\bibfnamefont {N.}~\bibnamefont {Mingo}},\ }\href {\doibase
  10.1080/21663831.2016.1174163} {\bibfield  {journal} {\bibinfo  {journal}
  {Materials Research Letters}\ }\textbf {\bibinfo {volume} {4}},\ \bibinfo
  {pages} {204} (\bibinfo {year} {2016})}\BibitemShut {NoStop}%
\bibitem [{\citenamefont {Serrano}\ \emph {et~al.}(2007)\citenamefont
  {Serrano}, \citenamefont {Bosak}, \citenamefont {Arenal}, \citenamefont
  {Krisch}, \citenamefont {Watanabe}, \citenamefont {Taniguchi}, \citenamefont
  {Kanda}, \citenamefont {Rubio},\ and\ \citenamefont {Wirtz}}]{SerBosAre07}%
  \BibitemOpen
  \bibfield  {author} {\bibinfo {author} {\bibfnamefont {J.}~\bibnamefont
  {Serrano}}, \bibinfo {author} {\bibfnamefont {A.}~\bibnamefont {Bosak}},
  \bibinfo {author} {\bibfnamefont {R.}~\bibnamefont {Arenal}}, \bibinfo
  {author} {\bibfnamefont {M.}~\bibnamefont {Krisch}}, \bibinfo {author}
  {\bibfnamefont {K.}~\bibnamefont {Watanabe}}, \bibinfo {author}
  {\bibfnamefont {T.}~\bibnamefont {Taniguchi}}, \bibinfo {author}
  {\bibfnamefont {H.}~\bibnamefont {Kanda}}, \bibinfo {author} {\bibfnamefont
  {A.}~\bibnamefont {Rubio}}, \ and\ \bibinfo {author} {\bibfnamefont
  {L.}~\bibnamefont {Wirtz}},\ }\href {\doibase 10.1103/PhysRevLett.98.095503}
  {\bibfield  {journal} {\bibinfo  {journal} {Physical Review Letters}\
  }\textbf {\bibinfo {volume} {98}},\ \bibinfo {pages} {095503} (\bibinfo
  {year} {2007})}\BibitemShut {NoStop}%
\bibitem [{\citenamefont {Tohei}\ \emph {et~al.}(2006)\citenamefont {Tohei},
  \citenamefont {Kuwabara}, \citenamefont {Oba},\ and\ \citenamefont
  {Tanaka}}]{TohKuwOba06}%
  \BibitemOpen
  \bibfield  {author} {\bibinfo {author} {\bibfnamefont {T.}~\bibnamefont
  {Tohei}}, \bibinfo {author} {\bibfnamefont {A.}~\bibnamefont {Kuwabara}},
  \bibinfo {author} {\bibfnamefont {F.}~\bibnamefont {Oba}}, \ and\ \bibinfo
  {author} {\bibfnamefont {I.}~\bibnamefont {Tanaka}},\ }\href {\doibase
  10.1103/PhysRevB.73.064304} {\bibfield  {journal} {\bibinfo  {journal}
  {Physical Review B}\ }\textbf {\bibinfo {volume} {73}},\ \bibinfo {pages}
  {064304} (\bibinfo {year} {2006})}\BibitemShut {NoStop}%
\bibitem [{\citenamefont {Reimers}\ \emph {et~al.}(2018)\citenamefont
  {Reimers}, \citenamefont {Sajid}, \citenamefont {Kobayashi},\ and\
  \citenamefont {Ford}}]{ReiSajKob18}%
  \BibitemOpen
  \bibfield  {author} {\bibinfo {author} {\bibfnamefont {J.~R.}\ \bibnamefont
  {Reimers}}, \bibinfo {author} {\bibfnamefont {A.}~\bibnamefont {Sajid}},
  \bibinfo {author} {\bibfnamefont {R.}~\bibnamefont {Kobayashi}}, \ and\
  \bibinfo {author} {\bibfnamefont {M.~J.}\ \bibnamefont {Ford}},\ }\href
  {\doibase 10.1021/acs.jctc.7b01072} {\bibfield  {journal} {\bibinfo
  {journal} {Journal of Chemical Theory and Computation}\ }\textbf {\bibinfo
  {volume} {14}},\ \bibinfo {pages} {1602} (\bibinfo {year}
  {2018})}\BibitemShut {NoStop}%
\bibitem [{\citenamefont {Abdi}\ \emph {et~al.}(2018)\citenamefont {Abdi},
  \citenamefont {Chou}, \citenamefont {Gali},\ and\ \citenamefont
  {Plenio}}]{AbdChoGal18}%
  \BibitemOpen
  \bibfield  {author} {\bibinfo {author} {\bibfnamefont {M.}~\bibnamefont
  {Abdi}}, \bibinfo {author} {\bibfnamefont {J.-P.}\ \bibnamefont {Chou}},
  \bibinfo {author} {\bibfnamefont {A.}~\bibnamefont {Gali}}, \ and\ \bibinfo
  {author} {\bibfnamefont {M.~B.}\ \bibnamefont {Plenio}},\ }\href {\doibase
  10.1021/acsphotonics.7b01442} {\bibfield  {journal} {\bibinfo  {journal} {ACS
  Photonics}\ }\textbf {\bibinfo {volume} {5}},\ \bibinfo {pages} {1967}
  (\bibinfo {year} {2018})}\BibitemShut {NoStop}%
\bibitem [{\citenamefont {Uddin}\ \emph {et~al.}(2017)\citenamefont {Uddin},
  \citenamefont {Li}, \citenamefont {Lin},\ and\ \citenamefont
  {Jiang}}]{UddLiLin17}%
  \BibitemOpen
  \bibfield  {author} {\bibinfo {author} {\bibfnamefont {M.~R.}\ \bibnamefont
  {Uddin}}, \bibinfo {author} {\bibfnamefont {J.}~\bibnamefont {Li}}, \bibinfo
  {author} {\bibfnamefont {J.~Y.}\ \bibnamefont {Lin}}, \ and\ \bibinfo
  {author} {\bibfnamefont {H.~X.}\ \bibnamefont {Jiang}},\ }\href {\doibase
  10.1063/1.4982647} {\bibfield  {journal} {\bibinfo  {journal} {Applied
  Physics Letters}\ }\textbf {\bibinfo {volume} {110}},\ \bibinfo {pages}
  {182107} (\bibinfo {year} {2017})}\BibitemShut {NoStop}%
\bibitem [{\citenamefont {Sajid}\ \emph {et~al.}(2018)\citenamefont {Sajid},
  \citenamefont {Reimers},\ and\ \citenamefont {Ford}}]{SajReiFor18}%
  \BibitemOpen
  \bibfield  {author} {\bibinfo {author} {\bibfnamefont {A.}~\bibnamefont
  {Sajid}}, \bibinfo {author} {\bibfnamefont {J.~R.}\ \bibnamefont {Reimers}},
  \ and\ \bibinfo {author} {\bibfnamefont {M.~J.}\ \bibnamefont {Ford}},\
  }\href {\doibase 10.1103/PhysRevB.97.064101} {\bibfield  {journal} {\bibinfo
  {journal} {Physical Review B}\ }\textbf {\bibinfo {volume} {97}},\ \bibinfo
  {pages} {064101} (\bibinfo {year} {2018})}\BibitemShut {NoStop}%
\bibitem [{\citenamefont {Wu}\ \emph {et~al.}(2017)\citenamefont {Wu},
  \citenamefont {Galatas}, \citenamefont {Sundararaman}, \citenamefont
  {Rocca},\ and\ \citenamefont {Ping}}]{WuGalSun17}%
  \BibitemOpen
  \bibfield  {author} {\bibinfo {author} {\bibfnamefont {F.}~\bibnamefont
  {Wu}}, \bibinfo {author} {\bibfnamefont {A.}~\bibnamefont {Galatas}},
  \bibinfo {author} {\bibfnamefont {R.}~\bibnamefont {Sundararaman}}, \bibinfo
  {author} {\bibfnamefont {D.}~\bibnamefont {Rocca}}, \ and\ \bibinfo {author}
  {\bibfnamefont {Y.}~\bibnamefont {Ping}},\ }\href {\doibase
  10.1103/PhysRevMaterials.1.071001} {\bibfield  {journal} {\bibinfo  {journal}
  {Physical Review Materials}\ }\textbf {\bibinfo {volume} {1}},\ \bibinfo
  {pages} {071001} (\bibinfo {year} {2017})}\BibitemShut {NoStop}%
\bibitem [{\citenamefont {Noh}\ \emph {et~al.}(2018)\citenamefont {Noh},
  \citenamefont {Choi}, \citenamefont {Kim}, \citenamefont {Im}, \citenamefont
  {Kim}, \citenamefont {Seo},\ and\ \citenamefont {Lee}}]{NohChoKim18}%
  \BibitemOpen
  \bibfield  {author} {\bibinfo {author} {\bibfnamefont {G.}~\bibnamefont
  {Noh}}, \bibinfo {author} {\bibfnamefont {D.}~\bibnamefont {Choi}}, \bibinfo
  {author} {\bibfnamefont {J.-H.}\ \bibnamefont {Kim}}, \bibinfo {author}
  {\bibfnamefont {D.-G.}\ \bibnamefont {Im}}, \bibinfo {author} {\bibfnamefont
  {Y.-H.}\ \bibnamefont {Kim}}, \bibinfo {author} {\bibfnamefont
  {H.}~\bibnamefont {Seo}}, \ and\ \bibinfo {author} {\bibfnamefont
  {J.}~\bibnamefont {Lee}},\ }\href {\doibase 10.1021/acs.nanolett.8b01030}
  {\bibfield  {journal} {\bibinfo  {journal} {Nano Letters}\ }\textbf {\bibinfo
  {volume} {18}},\ \bibinfo {pages} {4710} (\bibinfo {year}
  {2018})}\BibitemShut {NoStop}%
\bibitem [{\citenamefont {Kozawa}\ \emph {et~al.}(2019)\citenamefont {Kozawa},
  \citenamefont {Rajan}, \citenamefont {Li}, \citenamefont {Ichihara},
  \citenamefont {Koman}, \citenamefont {Zeng}, \citenamefont {Kuehne},
  \citenamefont {Iyemperumal}, \citenamefont {Silmore}, \citenamefont {Parviz},
  \citenamefont {Liu}, \citenamefont {Liu}, \citenamefont {Faucher},
  \citenamefont {Yuan}, \citenamefont {Xu}, \citenamefont {Warner},
  \citenamefont {Blankschtein},\ and\ \citenamefont {Strano}}]{KozGovXin19}%
  \BibitemOpen
  \bibfield  {author} {\bibinfo {author} {\bibfnamefont {D.}~\bibnamefont
  {Kozawa}}, \bibinfo {author} {\bibfnamefont {A.~G.}\ \bibnamefont {Rajan}},
  \bibinfo {author} {\bibfnamefont {S.~X.}\ \bibnamefont {Li}}, \bibinfo
  {author} {\bibfnamefont {T.}~\bibnamefont {Ichihara}}, \bibinfo {author}
  {\bibfnamefont {V.~B.}\ \bibnamefont {Koman}}, \bibinfo {author}
  {\bibfnamefont {Y.}~\bibnamefont {Zeng}}, \bibinfo {author} {\bibfnamefont
  {M.}~\bibnamefont {Kuehne}}, \bibinfo {author} {\bibfnamefont {S.~K.}\
  \bibnamefont {Iyemperumal}}, \bibinfo {author} {\bibfnamefont {K.~S.}\
  \bibnamefont {Silmore}}, \bibinfo {author} {\bibfnamefont {D.}~\bibnamefont
  {Parviz}}, \bibinfo {author} {\bibfnamefont {P.}~\bibnamefont {Liu}},
  \bibinfo {author} {\bibfnamefont {A.~T.}\ \bibnamefont {Liu}}, \bibinfo
  {author} {\bibfnamefont {S.}~\bibnamefont {Faucher}}, \bibinfo {author}
  {\bibfnamefont {Z.}~\bibnamefont {Yuan}}, \bibinfo {author} {\bibfnamefont
  {W.}~\bibnamefont {Xu}}, \bibinfo {author} {\bibfnamefont {J.~H.}\
  \bibnamefont {Warner}}, \bibinfo {author} {\bibfnamefont {D.}~\bibnamefont
  {Blankschtein}}, \ and\ \bibinfo {author} {\bibfnamefont {M.~S.}\
  \bibnamefont {Strano}},\ }\href {https://arxiv.org/abs/1909.11738} {\bibfield
   {journal} {\bibinfo  {journal} {arXiv:1909.11738}\ } (\bibinfo {year}
  {2019})}\BibitemShut {NoStop}%
\bibitem [{\citenamefont {Mackoit-Sinkevi\v{c}ien\.{e}}\ \emph
  {et~al.}(2019)\citenamefont {Mackoit-Sinkevi\v{c}ien\.{e}}, \citenamefont
  {Maciaszek}, \citenamefont {Van~de Walle},\ and\ \citenamefont
  {Alkauskas}}]{MacMacWal19}%
  \BibitemOpen
  \bibfield  {author} {\bibinfo {author} {\bibfnamefont {M.}~\bibnamefont
  {Mackoit-Sinkevi\v{c}ien\.{e}}}, \bibinfo {author} {\bibfnamefont
  {M.}~\bibnamefont {Maciaszek}}, \bibinfo {author} {\bibfnamefont {C.~G.}\
  \bibnamefont {Van~de Walle}}, \ and\ \bibinfo {author} {\bibfnamefont
  {A.}~\bibnamefont {Alkauskas}},\ }\href {\doibase 10.1063/1.5124153}
  {\bibfield  {journal} {\bibinfo  {journal} {Applied Physics Letters}\
  }\textbf {\bibinfo {volume} {115}},\ \bibinfo {pages} {212101} (\bibinfo
  {year} {2019})}\BibitemShut {NoStop}%
\bibitem [{\citenamefont {Berzina}\ \emph {et~al.}(2016)\citenamefont
  {Berzina}, \citenamefont {Korsaks}, \citenamefont {Trinkler}, \citenamefont
  {Sarakovskis}, \citenamefont {Grube},\ and\ \citenamefont
  {Bellucci}}]{BerKorTri16}%
  \BibitemOpen
  \bibfield  {author} {\bibinfo {author} {\bibfnamefont {B.}~\bibnamefont
  {Berzina}}, \bibinfo {author} {\bibfnamefont {V.}~\bibnamefont {Korsaks}},
  \bibinfo {author} {\bibfnamefont {L.}~\bibnamefont {Trinkler}}, \bibinfo
  {author} {\bibfnamefont {A.}~\bibnamefont {Sarakovskis}}, \bibinfo {author}
  {\bibfnamefont {J.}~\bibnamefont {Grube}}, \ and\ \bibinfo {author}
  {\bibfnamefont {S.}~\bibnamefont {Bellucci}},\ }\href {\doibase
  https://doi.org/10.1016/j.diamond.2016.06.010} {\bibfield  {journal}
  {\bibinfo  {journal} {Diamond and Related Materials}\ }\textbf {\bibinfo
  {volume} {68}},\ \bibinfo {pages} {131 } (\bibinfo {year}
  {2016})}\BibitemShut {NoStop}%
\bibitem [{\citenamefont {Pelini}\ \emph {et~al.}(2019)\citenamefont {Pelini},
  \citenamefont {Elias}, \citenamefont {Page}, \citenamefont {Xue},
  \citenamefont {Liu}, \citenamefont {Li}, \citenamefont {Edgar}, \citenamefont
  {Dr\'eau}, \citenamefont {Jacques}, \citenamefont {Valvin}, \citenamefont
  {Gil},\ and\ \citenamefont {Cassabois}}]{PelEliPag19}%
  \BibitemOpen
  \bibfield  {author} {\bibinfo {author} {\bibfnamefont {T.}~\bibnamefont
  {Pelini}}, \bibinfo {author} {\bibfnamefont {C.}~\bibnamefont {Elias}},
  \bibinfo {author} {\bibfnamefont {R.}~\bibnamefont {Page}}, \bibinfo {author}
  {\bibfnamefont {L.}~\bibnamefont {Xue}}, \bibinfo {author} {\bibfnamefont
  {S.}~\bibnamefont {Liu}}, \bibinfo {author} {\bibfnamefont {J.}~\bibnamefont
  {Li}}, \bibinfo {author} {\bibfnamefont {J.~H.}\ \bibnamefont {Edgar}},
  \bibinfo {author} {\bibfnamefont {A.}~\bibnamefont {Dr\'eau}}, \bibinfo
  {author} {\bibfnamefont {V.}~\bibnamefont {Jacques}}, \bibinfo {author}
  {\bibfnamefont {P.}~\bibnamefont {Valvin}}, \bibinfo {author} {\bibfnamefont
  {B.}~\bibnamefont {Gil}}, \ and\ \bibinfo {author} {\bibfnamefont
  {G.}~\bibnamefont {Cassabois}},\ }\href {\doibase
  10.1103/PhysRevMaterials.3.094001} {\bibfield  {journal} {\bibinfo  {journal}
  {Physical Review Materials}\ }\textbf {\bibinfo {volume} {3}},\ \bibinfo
  {pages} {094001} (\bibinfo {year} {2019})}\BibitemShut {NoStop}%
\bibitem [{\citenamefont {Hamdi}\ \emph {et~al.}(2020)\citenamefont {Hamdi},
  \citenamefont {Thiering}, \citenamefont {Bodrog}, \citenamefont {Ivády},\
  and\ \citenamefont {Gali}}]{HamThiBod20}%
  \BibitemOpen
  \bibfield  {author} {\bibinfo {author} {\bibfnamefont {H.}~\bibnamefont
  {Hamdi}}, \bibinfo {author} {\bibfnamefont {G.}~\bibnamefont {Thiering}},
  \bibinfo {author} {\bibfnamefont {Z.}~\bibnamefont {Bodrog}}, \bibinfo
  {author} {\bibfnamefont {V.}~\bibnamefont {Ivády}}, \ and\ \bibinfo {author}
  {\bibfnamefont {A.}~\bibnamefont {Gali}},\ }\href {\doibase
  10.1038/s41524-020-00451-y} {\bibfield  {journal} {\bibinfo  {journal} {npj
  Computational Materials}\ }\textbf {\bibinfo {volume} {6}},\ \bibinfo {pages}
  {1} (\bibinfo {year} {2020})}\BibitemShut {NoStop}%
\bibitem [{\citenamefont {Vokhmintsev}\ and\ \citenamefont
  {Weinstein}(2020)}]{VokWei20}%
  \BibitemOpen
  \bibfield  {author} {\bibinfo {author} {\bibfnamefont {A.~S.}\ \bibnamefont
  {Vokhmintsev}}\ and\ \bibinfo {author} {\bibfnamefont {I.~A.}\ \bibnamefont
  {Weinstein}},\ }\href@noop {} {\bibfield  {journal} {\bibinfo  {journal}
  {arXiv:2003.02789}\ } (\bibinfo {year} {2020})}\BibitemShut {NoStop}%
\bibitem [{\citenamefont {Sohier}\ \emph {et~al.}(2017)\citenamefont {Sohier},
  \citenamefont {Gibertini}, \citenamefont {Calandra}, \citenamefont {Mauri},\
  and\ \citenamefont {Marzari}}]{SohGibCal17}%
  \BibitemOpen
  \bibfield  {author} {\bibinfo {author} {\bibfnamefont {T.}~\bibnamefont
  {Sohier}}, \bibinfo {author} {\bibfnamefont {M.}~\bibnamefont {Gibertini}},
  \bibinfo {author} {\bibfnamefont {M.}~\bibnamefont {Calandra}}, \bibinfo
  {author} {\bibfnamefont {F.}~\bibnamefont {Mauri}}, \ and\ \bibinfo {author}
  {\bibfnamefont {N.}~\bibnamefont {Marzari}},\ }\href {\doibase
  10.1021/acs.nanolett.7b01090} {\bibfield  {journal} {\bibinfo  {journal}
  {Nano Letters}\ }\textbf {\bibinfo {volume} {17}},\ \bibinfo {pages} {3758}
  (\bibinfo {year} {2017})}\BibitemShut {NoStop}%
\end{thebibliography}
\end{document}